# Muon momentum measurement in ICARUS-T600 LAr-TPC via multiple scattering in few-GeV range

**M. Antonello[a], B. Baibussinov[b], V. Bellini[c], P. Benetti[d], F. Boffelli[d], A. Bubak[e], E. Calligarich[d], S. Centro[b], T. Cervi[d], A. Cesana[f], K. Cieslik[g], A.G. Cocco[h], A. Dabrowska[g], A. Dermenev[i], A. Falcone[d,†], C. Farnese[b], A. Fava[b,‡], A. Ferrari[j], D. Gibin[b], S. Gninenko[i], A. Guglielmi[b], M. Haranczyk[g], J. Holeczek[e], M. Janik[e], M. Kirsanov[i], J. Kisiel[e], I. Kochanek[e], J. Lagoda[k], A. Menegolli[d], G. Meng[b], C. Montanari[d], S. Otwinowski[l], P. Picchi[m], F. Pietropaolo[b], P. Plonski[n], A. Rappoldi[d], G.L. Raselli[d], M. Rossella[d], C. Rubbia[a,j,o], P. Sala[f], A. Scaramelli[d], F. Sergiampietri[p], M. Spanu[d], D. Stefan[f], R. Sulej[e], M. Szarska[g], M. Terrani[f], M. Torti[d], F. Tortorici[c], F. Varanini[b,§], S. Ventura[b], C. Vignoli[a], H. Wang[l], X. Yang[l], A. Zalewska[g], A. Zani[d], K. Zaremba[n]**

**(ICARUS Collaboration)**

[a] *INFN - Laboratori Nazionali del Gran Sasso, Assergi, Italy*
[b] *Dipartimento di Fisica e Astronomia Università di Padova and INFN, Padova, Italy*
[c] *Dipartimento di Fisica e Astronomia Università di Catania and INFN, Catania, Italy*
[d] *Dipartimento di Fisica Università di Pavia and INFN, Pavia, Italy*
[e] *Institute of Physics, University of Silesia, Katowice, Poland*
[f] *INFN, Milano, Italy*
[g] *H. Niewodniczanski Institute of Nuclear Physics, Polish Academy of Science, Krakow, Poland*
[h] *Dipartimento di Scienze Fisiche Università Federico II di Napoli and INFN, Napoli, Italy*
[i] *INR RAS, Moscow, Russia*
[j] *CERN, Geneva, Switzerland*
[k] *National Centre for Nuclear Research, Otwock/Swierk, Poland*
[l] *Department of Physics and Astronomy, UCLA, Los Angeles, USA*
[m] *INFN Laboratori Nazionali di Frascati, Frascati, Italy*
[n] *Institute of Radioelectronics, Warsaw University of Technology, Warsaw, Poland*
[o] *GSSI, L'Aquila, Italy*
[p] *INFN, Pisa, Italy*

*E-mail*: filippo.varanini@pd.infn.it

[†] Now at University of Texas, Arlington, USA
[‡] Now at Fermilab
[§] Corresponding author

ABSTRACT: The measurement of muon momentum by Multiple Coulomb Scattering is a crucial ingredient to the reconstruction of $\nu_\mu$ CC events in the ICARUS-T600 liquid argon TPC in absence of magnetic field, as in the search for sterile neutrinos at Fermilab where ICARUS will be exposed to ~1 GeV Booster neutrino beam. A sample of ~1000 stopping muons produced by charged current interactions of CNGS $\nu_\mu$ in the surrounding rock at the INFN Gran Sasso underground Laboratory provides an ideal benchmark in the few-GeV range since their momentum can be directly and independently obtained by the calorimetric measurement. Stopping muon momentum in the 0.5- 4.5 GeV/c range has been reconstructed via Multiple Coulomb Scattering with resolution ranging from 10 to 25 % depending on muon energy, track length and uniformity of the electric field in the drift volume.

## Contents

---

## 1. Introduction

The innovative Liquid Argon Time Projection Chamber (LAr-TPC) detection technique [1] is characterized by precise imaging and calorimetric reconstruction capabilities of any ionizing track in neutrino processes or other rare events with a performance comparable to a traditional bubble chamber. In addition the LAr-TPC is a fully electronic, continuously sensitive and self-triggering detector. The operating principle is based on the fact that in highly purified liquid Argon free electrons from ionizing particles can be easily transported over macroscopic distances (meters) with the help of a uniform electric field to a multi-wire anodic structure placed at the end of the drift path.

The ICARUS Collaboration developed the LAr-TPC technology from prototype dimensions to the ~ 1 kt mass of the T600 LAr detector [2] installed in the underground INFN-LNGS Gran Sasso Laboratory. The successful and extended operation has demonstrated the potential of this novel detection technique, while developing a vast physics program including the observation of neutrinos both from the CNGS beam at a distance of 730 km from CERN and from cosmic rays. ICARUS-T600 marks a major milestone in the development of large-scale LAr detectors as required for present short [3] and future long [4] base-line neutrino projects.

The T600 detector is presently under overhauling at CERN in view of the next exploitation at shallow depth at FNAL in the framework of the SBN [3] neutrino program searching for sterile neutrino where it will be soon exposed to ~1 GeV Booster neutrinos. Moreover ICARUS-T600 will receive also in off-axis position the NuMI neutrino beam, enriched in electron neutrinos, recording a large sample of neutrino interactions in few GeV energy range.

The energy of hadrons, electrons and γ's is reconstructed in the LAr-TPC by accurate calorimetric measurement of the related showers. However, this approach is not suitable for muons escaping the active volume, since they have no appreciable nuclear interactions and may travel over much longer distances (their minimum ionization density in LAr is ~0.21 GeV/m). Since no magnetic field is present in the ICARUS-T600 detector, Multiple Coulomb scattering (MCS) is the only alternative to estimate the momentum of muons escaping the detector. The momentum measurement by MCS is particularly useful in neutrino interactions in the one to few GeV energy range, such as in the Booster and NuMI neutrino beams, since in this case a substantial fraction of muons produced in the neutrino charged current interactions are emitted at large angles and escape the detector, preventing a calorimetric measurement of the energy.

The method of momentum measurement via MCS in a LAr-TPC proposed by C. Rubbia was applied to cosmic muon events recorded during the ICARUS-T600 technical run on Earth surface in Pavia [1]. The momentum of stopping muons below 0.8 GeV/c was previously measured via MCS and compared with the calorimetric measurement [5]. The present study extends the momentum measurement by MCS up to 5 GeV covering the energy range relevant for the SBN experiment [3] and for the next generation neutrino experiments [4]. The performance of the momentum measurement algorithm has been determined with a calibration sample of ~500 muons produced in interactions of CNGS neutrinos in the upstream rock at LNGS and stopping inside the T600. Muon momentum by MCS relies on the precise determination of the track hits position along the drift coordinate. The accuracy of the latter depends on the uniformity of the drift velocity and, as a consequence, of the drift electric field.

## 2. The ICARUS-T600 detector

The ICARUS-T600 detector consists of a large cryostat filled with about 760 t of ultra-pure liquid argon, split into two identical adjacent modules (identified as East and West modules) surrounded by a thermal insulation vessel [1][2]. A thermal shield is placed between the insulation and the Aluminum containers, with boiling Nitrogen circulating inside to intercept the heat load and maintain the cryostat bulk temperature uniform and stable at 89 K, by means of a dedicated re-liquefaction system of cryo-coolers. To keep electronegative impurities in LAr at a very low concentration level, each module is equipped with two gas and one liquid Argon recirculation/purification systems. The adopted solutions permitted to reach a LAr purity at the level of ~20 ppt $O_2$ equivalent, corresponding to free electron lifetimes exceeding 15 ms [6].

Each module houses two TPC chambers with 1.5 m maximum drift path, sharing a common central cathode, composed by 9 punched stainless-steel panels ~2 m long and ~3.2 m high, vertically installed within a metallic framework with a central horizontal reinforcing bar. The cathode plane and the field cage electrodes, composed by stainless-steel tubes (34 mm diameter and ~50 mm pitch, with a mechanical precision better than 0.5 mm), surrounding the LAr volume generate an ideally uniform electric field $E_{DRIFT}$ = 500 V/cm. Ionization electrons are then drifted with velocity $v_D$ ~1.6 mm/μs, toward three wire arrays installed on the sides, allowing a stereoscopic event reconstruction; the planarity of the wire chambers is by design better than 100 μm. A total of 53248 wires are deployed, with a 3 mm pitch, oriented on each plane at a different angle (0°, ±60°) with respect to the horizontal direction. A schematic picture of a module with the corresponding coordinates used in this analysis is shown in Figure 1.

By appropriate voltage biasing, the first two wire planes (Induction1 and Induction2) record signals in a non-destructive way, while the ionization charge is collected and measured on the last plane (Collection). The electronics was designed to allow continuous read-out, digitization and independent waveform recording of signals from each wire of the TPC. The read-out chain is organized on a 32-channel modularity. Signals of the charge sensitive front-end amplifiers have been digitized with 10-bits ADCs with $t_s$=400 ns sampling time. The overall gain is about 1000 electrons for each ADC count, setting the pulse height in the Collection view of a minimum ionizing particle (m.i.p.) signal to ~15 ADC counts for a 3 mm pitch. The corresponding average electronic noise has been estimated to ~1.5 ADC counts, and the gain uniformity has been measured by injecting test-pulse signals on each channel, with a relative accuracy better than 5%, determined by the uncertainties on the adopted calibration capacitances. The Induction signals are characterized by a smaller signal/noise ratio, with an amplitude of ~10 ADC counts in both views, for tracks parallel to the wire planes.

In order to determine the absolute position of the track along the drift coordinate, the electron arrival time to the wire chambers is compared with the prompt scintillation light in LAr recorded by the ICARUS photo-multiplier system installed behind the TPC wires [7].

The drift time interval is then converted into an absolute distance along the drift coordinate ($y$) assuming a nominal uniform drift velocity corresponding to the average measured value.

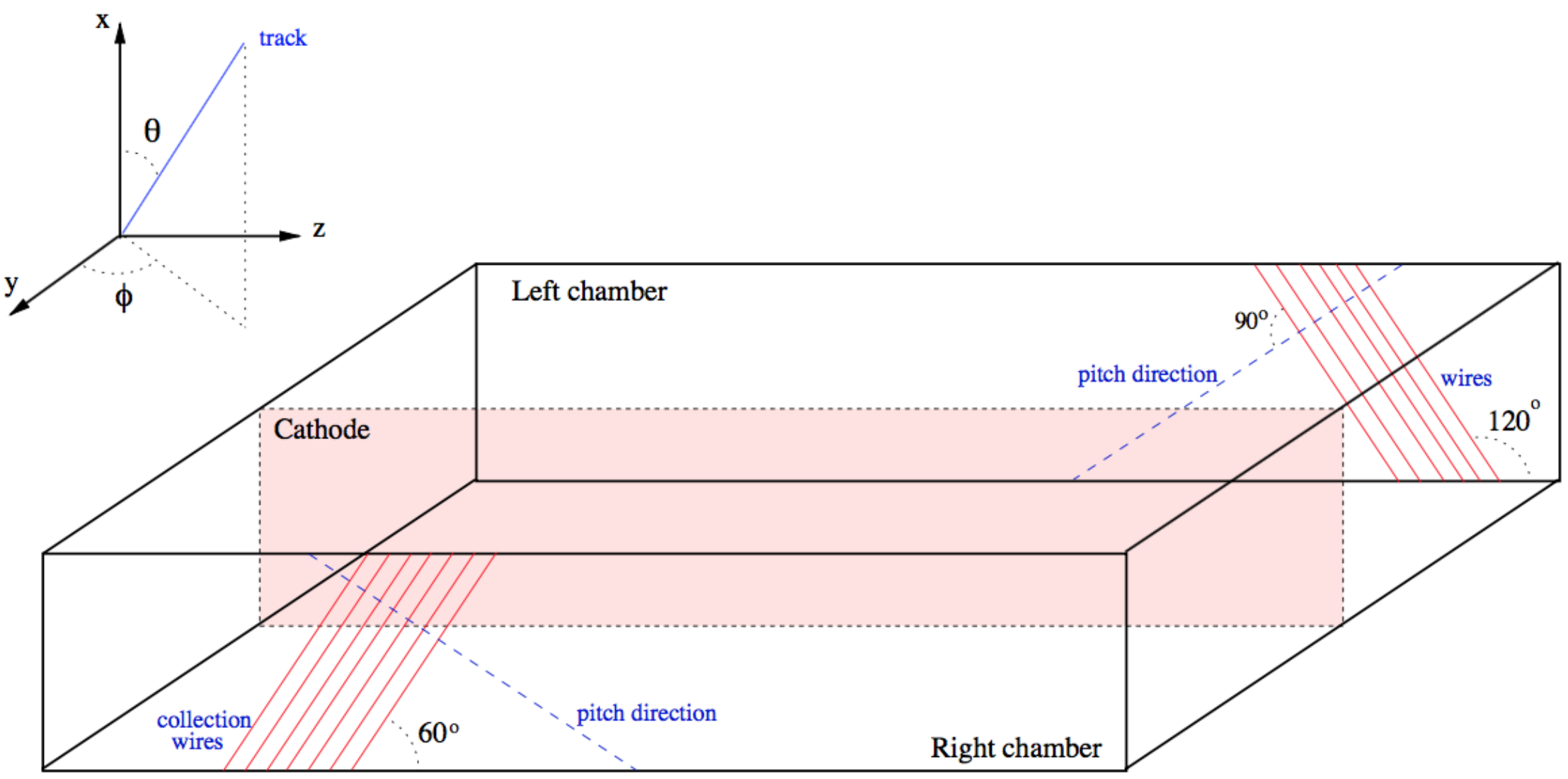

**Figure 1: Scheme of a T300 module, including Collection wire directions and the coordinate system used in this work: $z$ is longitudinal (along the beam direction), x is vertical, while y is along the drift direction.**

Local deviations from planarity in the central cathode up to 2.5 cm (Figure 2), extending over few meter lengths, have been measured with a laser-meter during an inspection at CERN of the TPC's of the East module after the decommissioning of the ICARUS detector. This deformation affects the electric field uniformity in the chamber and the reconstructed $y$ drift coordinate. An independent measurement of the cathode lack of planarity has been obtained with cosmic muon tracks traversing the cathode recorded during the data taking at LNGS. The difference between the maximum drift times of the tracks in the two adjacent chambers was used to estimate the cathode distortion at the crossing point between the chambers. The two estimations appear correlated for most of the cathode panels (Figure 3), despite referring to two vastly different experimental conditions, i.e. the LAr temperature versus room temperature, after the Gran Sasso to CERN module transport.

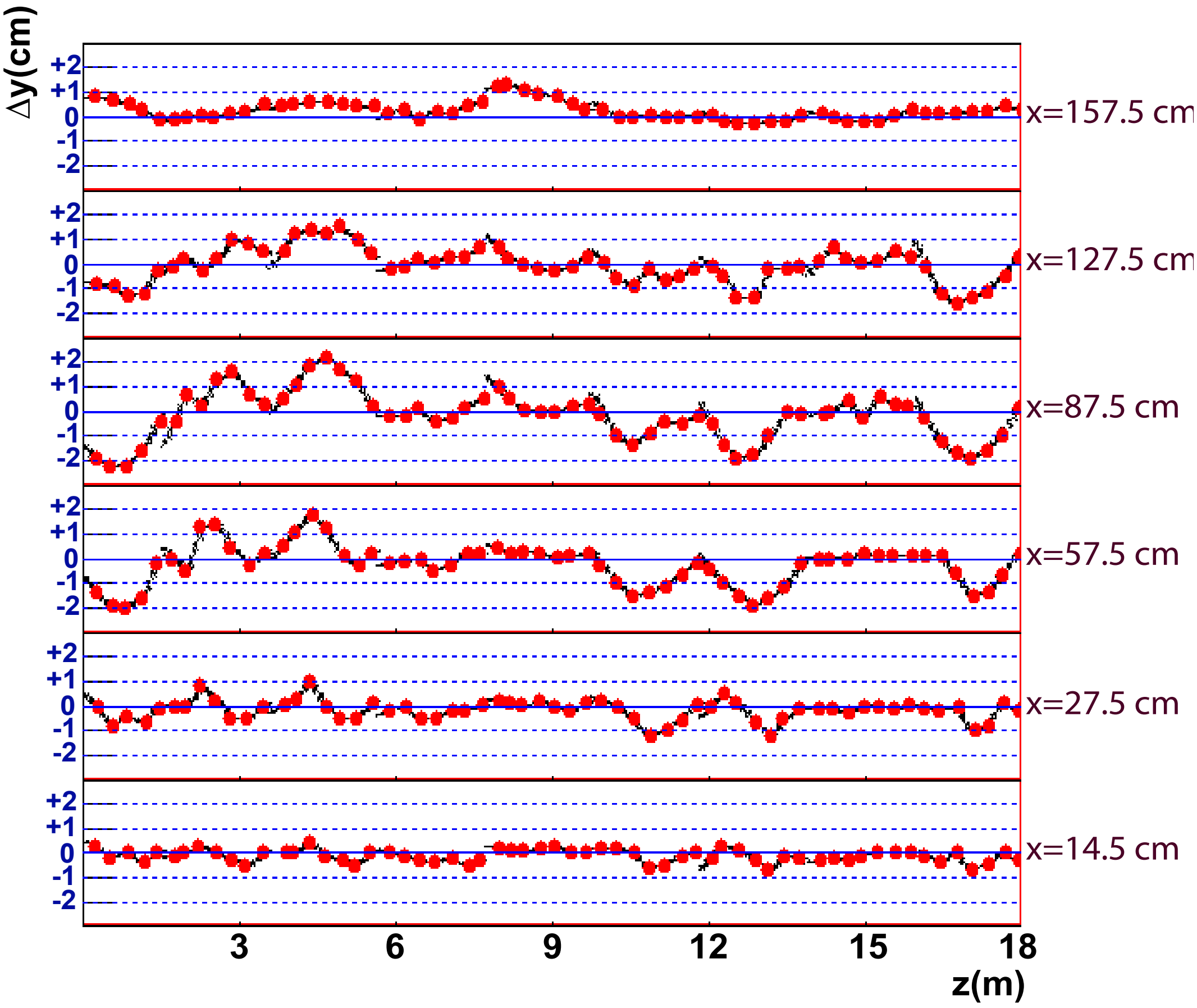

**Figure 2: Measured deviations from planarity of the lower panels of the cathode (red dots) along the drift coordinate (*y*) as a function of longitudinal coordinate (*z*), at various heights *x* (shown on the right side of the plot). Black points are a geometrical bilinear interpolation of the position between the measured values, blue solid lines refer to the nominal flat cathode position.**

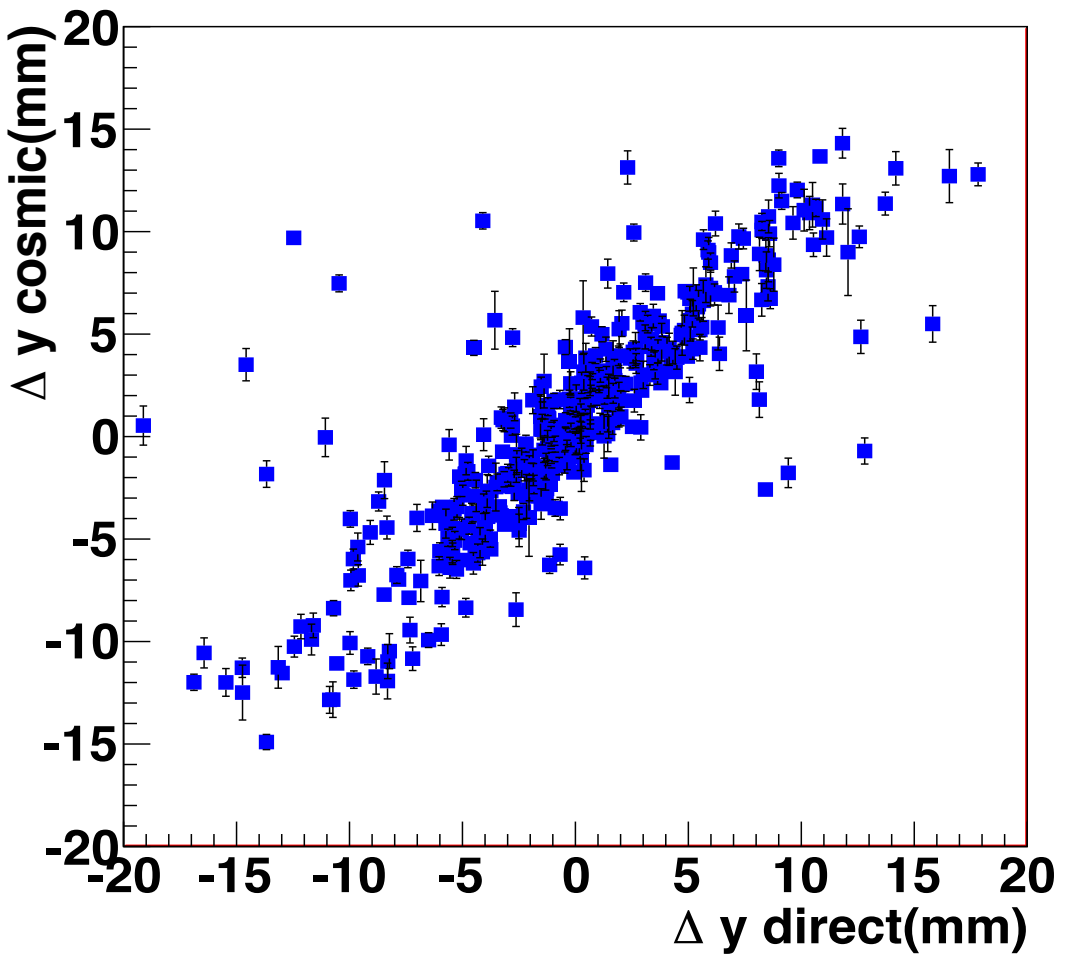

**Figure 3: Comparison of direct and cosmic muons estimations of cathode distortions. Each entry corresponds to the measured distortion averaged over a pixel of ~30x30 cm$^2$ area on the cathode surface (only statistical errors are quoted).**

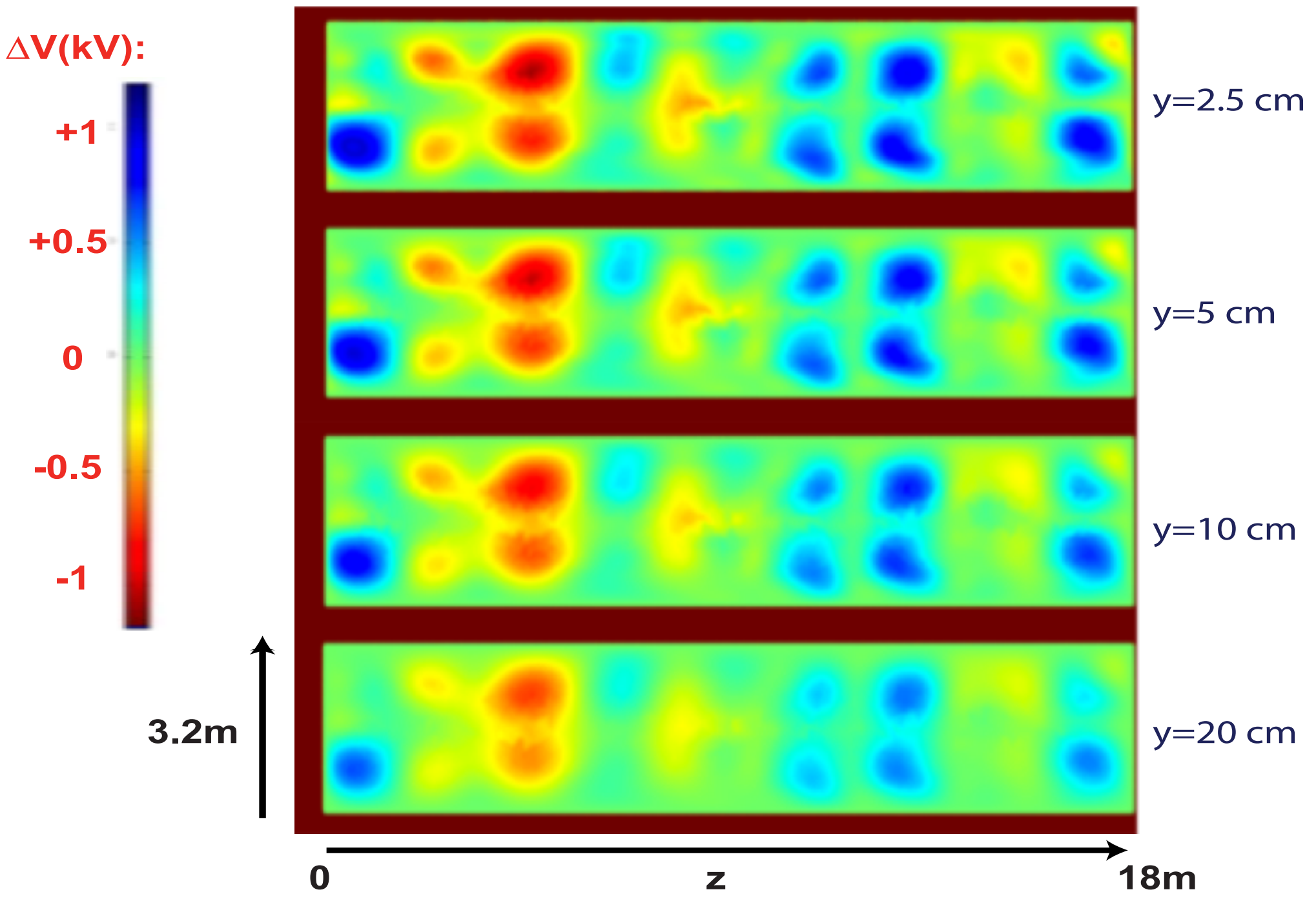

**Figure 4: Electric potential distortions from the nominal value obtained from a 3D field simulation, using measured cathode distortions as an input. Four vertical slices (3.2 m x 18 m), at y=2.5, 5, 10 and 20 cm distance from the cathode, are shown. The scale of the distortion runs from -1 kV (red) to 1 kV (blue). The cathode mechanical structure can be recognized: the 9 cathode panels with ~2 m length are clearly visible. Deformations are usually larger in the middle of each panel and vanish on the edges. They also tend to vanish at ~1.6 m height, near the horizontal reinforcing bar.**

A full 3D numerical calculation of the electric field on a grid of 2.5 cm mesh has been performed with the Comsol package [8] according to the laser-meter measurements (Figure 3). The calculated longitudinal profiles of the drift field at 30 cm distance from the cathode (Figure 4) follow the structure of the measured deviations.

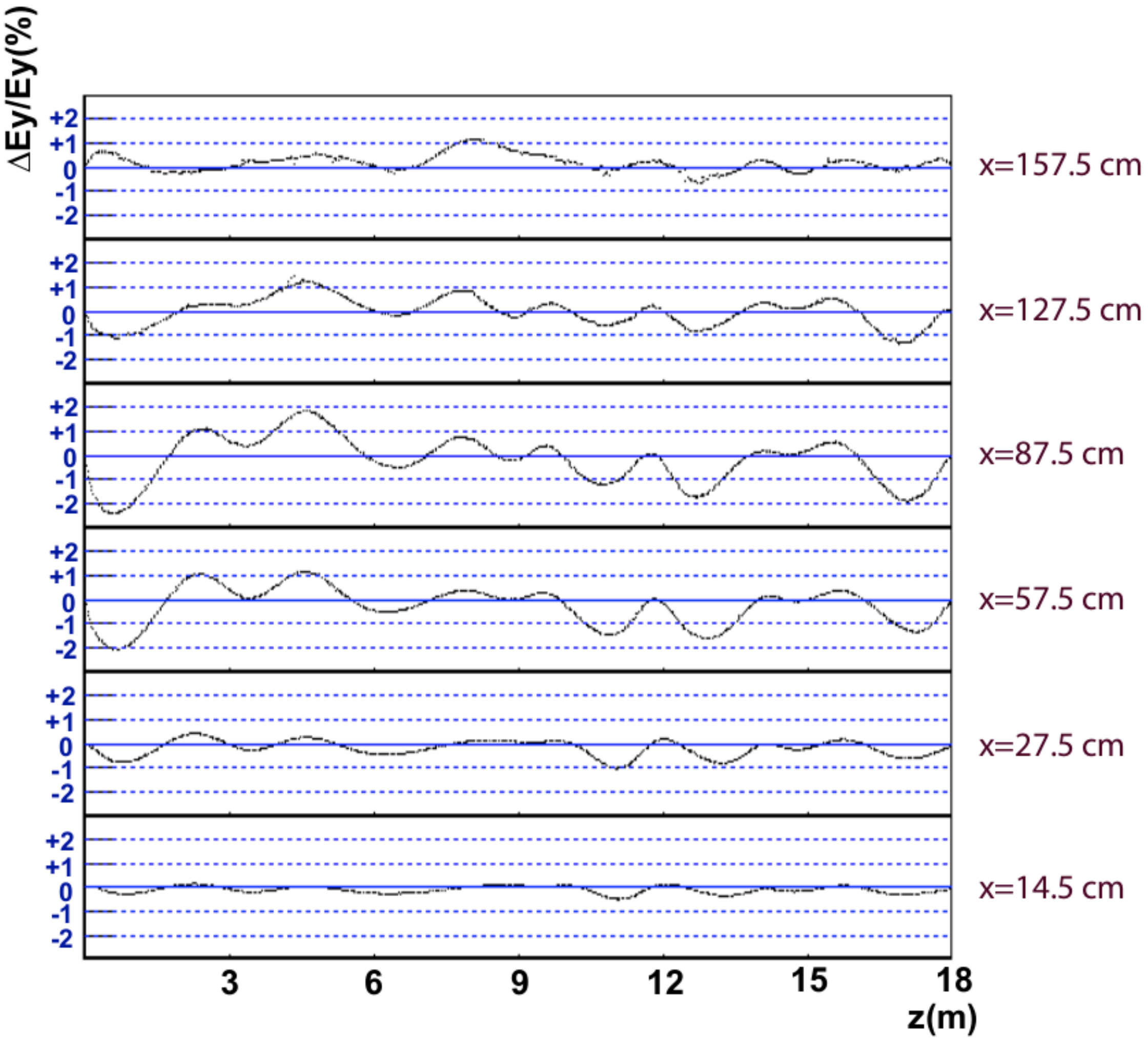

**Figure 5: Electric field distortion from 500 V/cm (solid lines) in the lower half of the cathode obtained from a 3D field simulation, using measured cathode deformations as an input. Field distortions are shown as functions of the longitudinal coordinate *z*, at 30 cm distance from the cathode in the left chamber, at various heights *x* (shown on the right side). Dashed lines correspond to ±1% and ±2% differences.**

A drift field distortion in excess of few percent, with an opposite sign in the two adjacent chambers, is observed in the simulation in the proximity of the cathode, decreasing roughly exponentially with the distance (Figure 6). The corresponding distortion $\delta y$ in the reconstructed drift coordinate has been computed by a numerical integration of the local effective electron drift velocity $v_D \sim E_y^{1/2}$ along the drift path. As a result a several mm spatial distortion has been determined for tracks running in the proximity of the cathode (Figure 7).

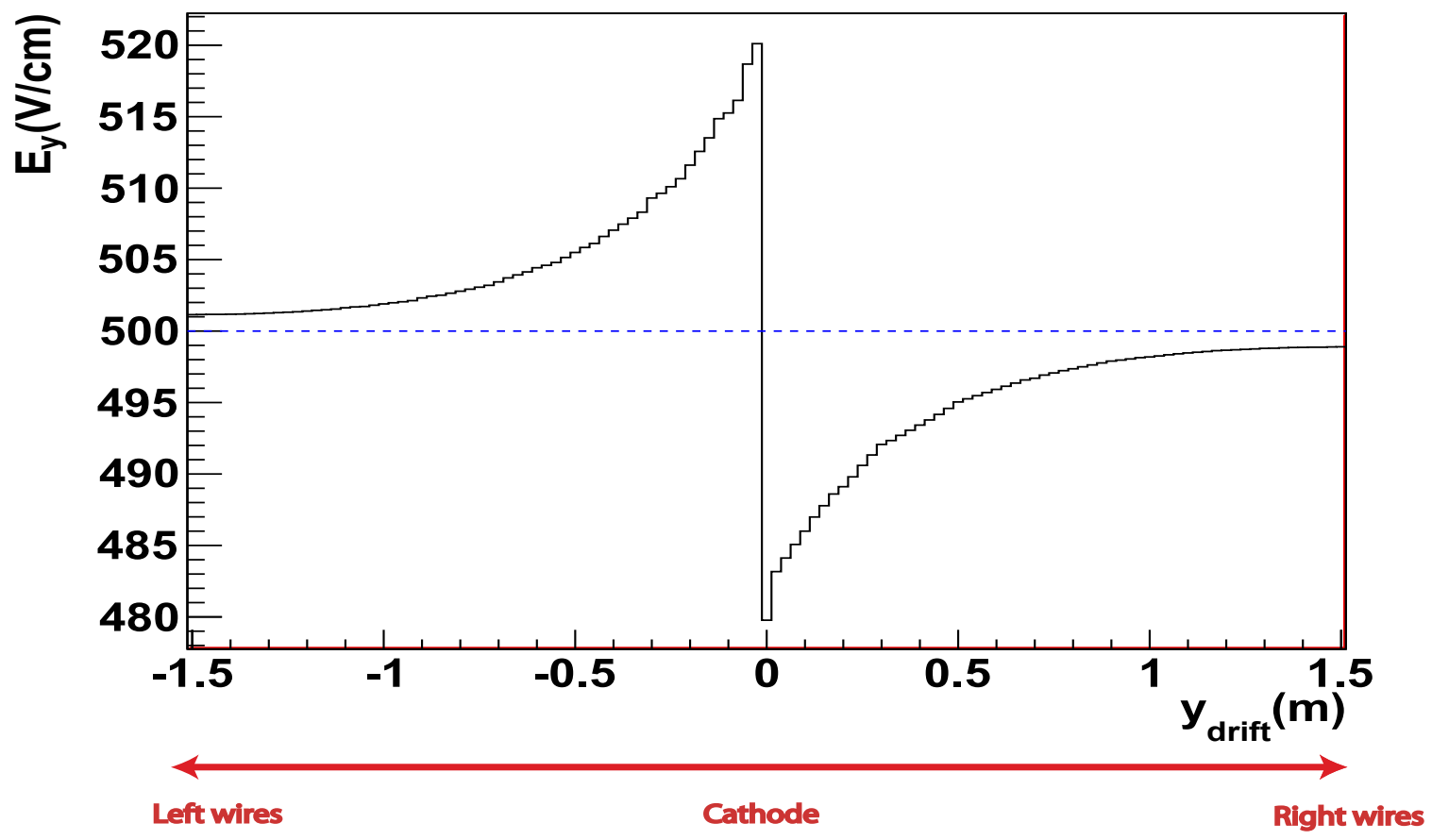

**Figure 6: Electric field distortions from 500 V/cm nominal value as a function of the drift coordinate in the middle of the first upstream lower panel, obtained from full 3D field simulation using measured cathode distortions as an input.**

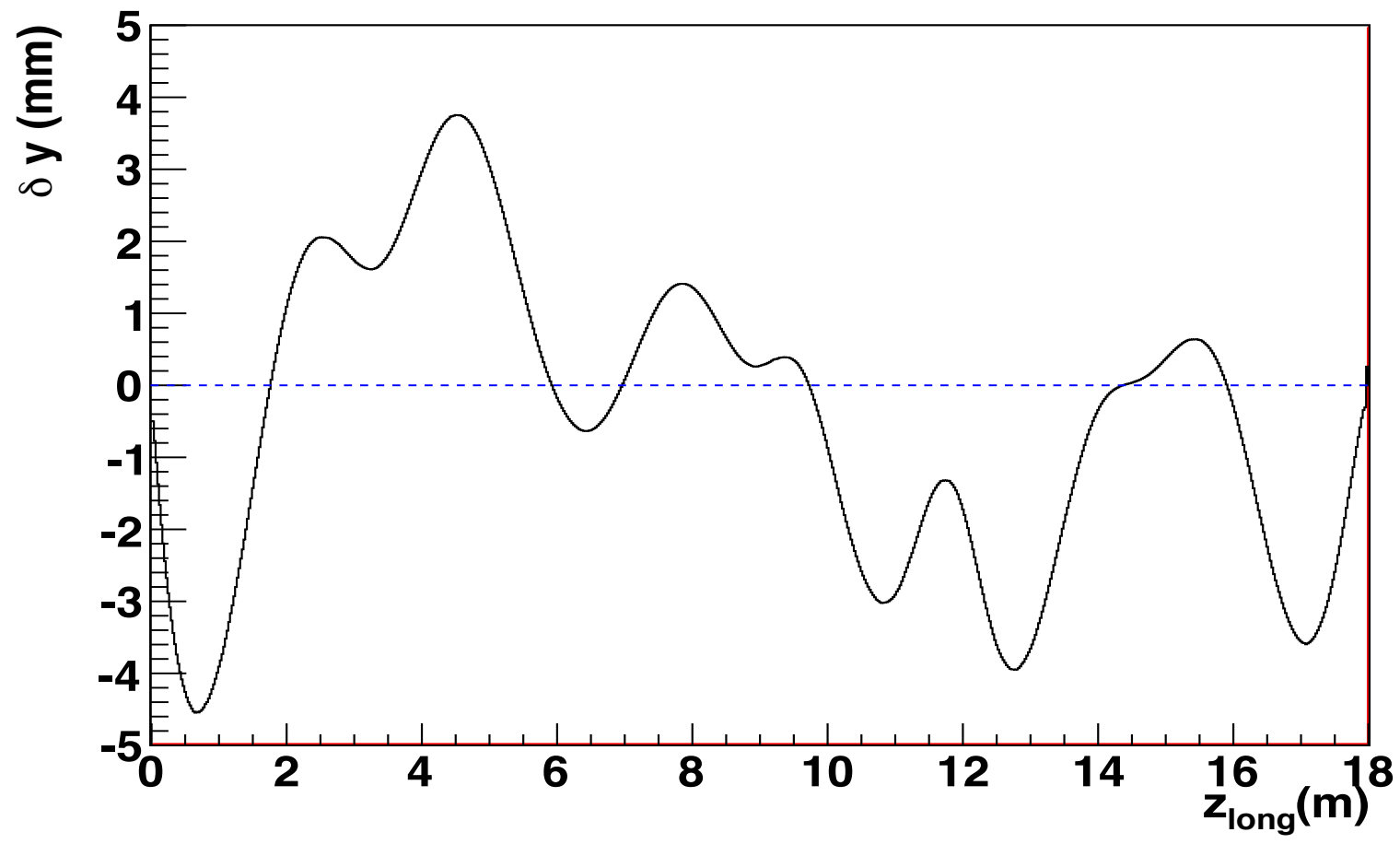

**Figure 7: Profile of the distortion δy along the longitudinal coordinate z at 30cm distance from cathode, obtained from full 3D field simulation using measured cathode distortions as an input.**

A mechanical intervention allowed to restore the cathode planarity within a few mm from the initial deviation of 2.5 cm by warming up and pressing the panels during the overhauling phase of the ICARUS-T600 detector in view of the SBN sterile neutrino program at FNAL Booster beam [3].

## 3. The ICARUS-T600 stopping muon sample

The present analysis refers to ~1000 visually selected stopping muon candidates, generated in the CNGS beam neutrino interactions in the T600 upstream rock, where hadrons are fully absorbed while the muon enters the ICARUS detector and stops in the LAr volume. A subsample of 460 stopping muon events has been identified, requiring a minimum track length of 2.5 m (~3 hadronic interaction lengths in LAr) without visible nuclear interactions along the track. The collected data correspond to a statistics of ~ 5x10$^{19}$ protons on target. The regions close to the detector borders (1 cm from wire planes and 5 cm from central cathode and other sides of the TPCs) have been excluded because of possible large distortions of the electric field, resulting in a fiducial LAr mass of 433 t.

The calorimetric momentum, $p_{CAL}$, of each muon has been computed from its kinetic energy ($E_k$) that was measured by integrating all the ionization charge collected along the muon track, including δ-rays and bremsstrahlung photons [9], The electron-ion recombination is corrected for, as explained in [10]: the typical fraction of electrons surviving recombination is ~0.69 for a m.i.p in LAr. The signal attenuation along the drift path due to electronegative impurities has also been taken into account event by event. LAr purity was constantly monitored during data-taking as described in [6] and the measured electron lifetime turned out to always exceed ~7 ms, corresponding to a maximum attenuation of ~12%.

An alternative, purely geometrical, estimation of muon momentum ($p_{RANGE}$), obtained from range, with a worse resolution (~4%, mostly due to straggling, to be compared with ~1.5% for calorimetry) but immune from calorimetric corrections, was used to check the $p_{CAL}$ measurement, showing that the average ratio $p_{CAL}/p_{RANGE}$ in data and MC agree to better than 2%. After correcting for this small difference, the accuracy of $p_{CAL}$ in data is estimated to be better than ~1%.

The kinetic energy spectrum of the selected stopping muons is shown in Figure 8, as well as the corresponding range.

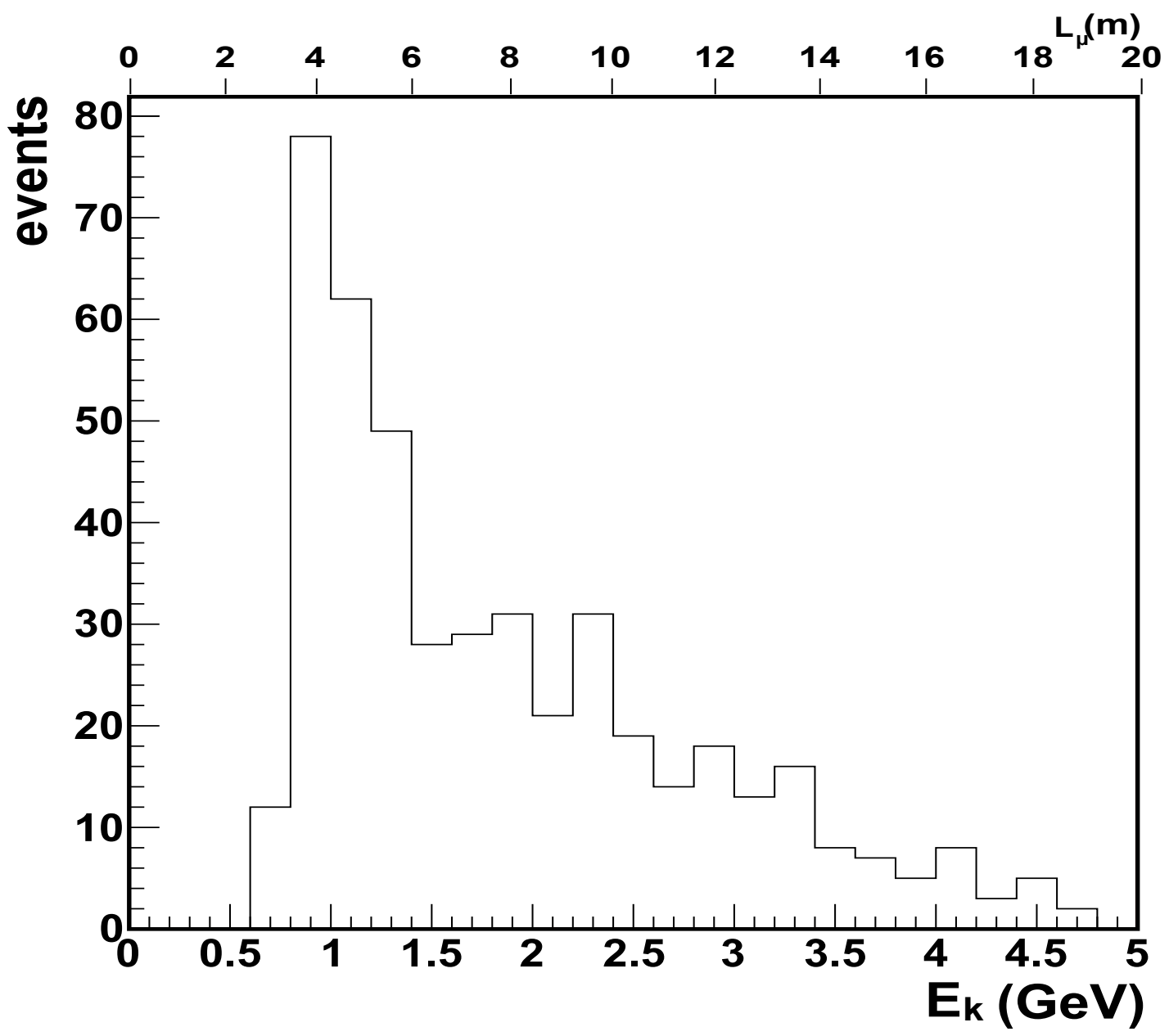

**Figure 8: Distribution of stopping muons as a function of their kinetic energy (bottom axis) and corresponding muon length (top axis).**

In order to evaluate the resolution of momentum measurement for non-contained tracks in real experiments, only a fixed initial length $L_\mu$ of the muon tracks has been considered; in the default case used for almost all results shown in this paper $L_\mu$ is 4 meters. Only muons longer than $L_\mu$+1 meter were studied in this analysis (corresponding to ~300 events in the default case), in order to avoid biasing the measurement with the knowledge of the higher ionization density in the final region of the muon track, close to range-out, where dE/dx increases significantly.

A track cleaning procedure has been applied to remove outlier hits, mainly produced by large δ-rays that may mimic significant track deflections and strongly distort the momentum measurement. Delta rays can be identified either as branches separated from the muon track, or as large energy depositions on multiple wires, and correspond to ~11% of hits in the stopping muon sample. A small fraction of residual outliers (~2%).have been removed by a $\chi^2$-based cut after a Kalman fit ([11],[12]) of the muon track . As an example, the effect of the cleaning procedure on a stopping muon is shown in Figure 9.

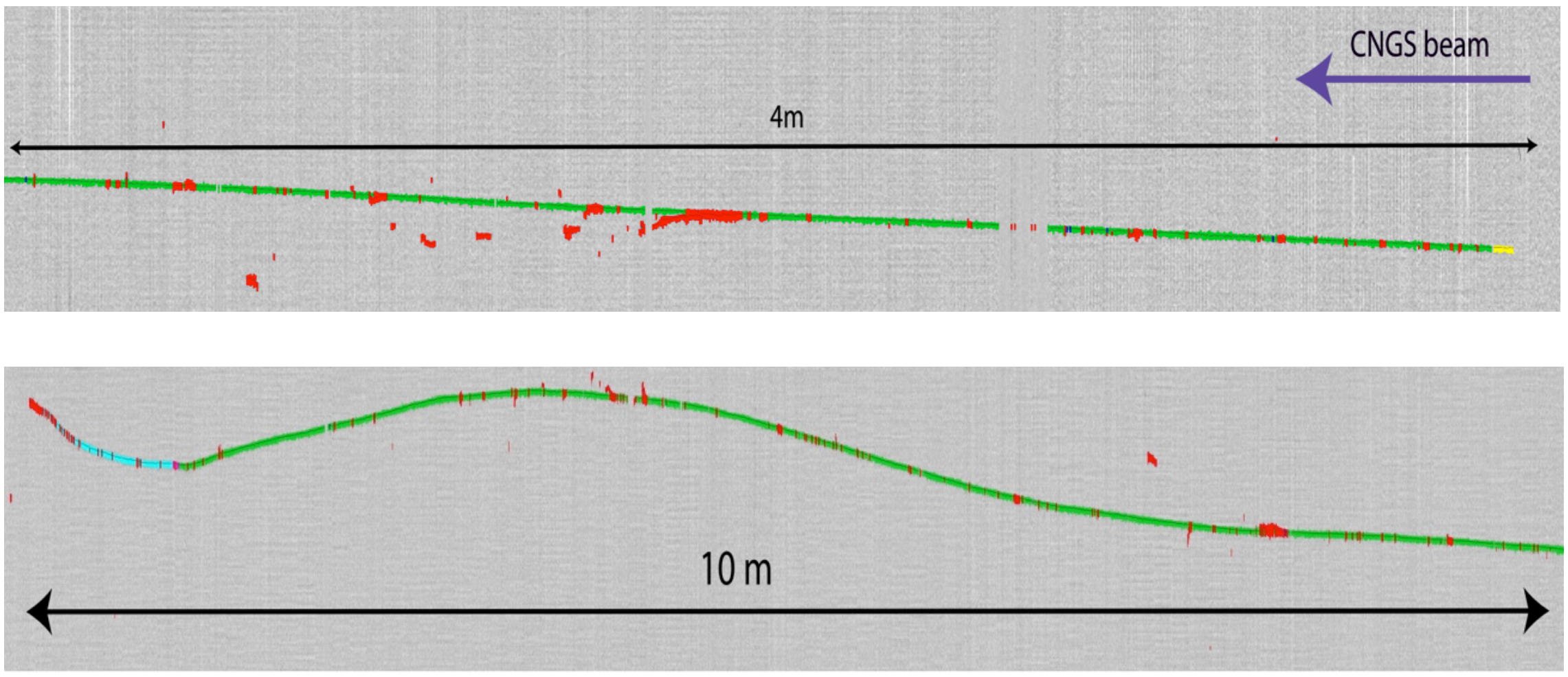

**Figure 9: Example of the cleaning procedure in a 14.6 m stopping μ track. The hits considered for MCS analysis are shown in green; the identified δ-rays and the associated γs (in red) have been excluded from the analysis. The first μ hits (in yellow) in the zoomed initial part of the track (top) are out of fiducial volume (see text). In the second part of the trajectory (bottom), hits in the final meter of the track (in light blue) are also excluded. The accuracy on the drift time on wires is a factor ≈10 better than the width of the colored band in the figure, representing the duration of the wire signal.**

## 4. Estimate of the measurement error on the drift coordinate

Measurement errors in determining the position of μ hits in the TPC result in apparent track deflections that can mimic the physical effect of multiple scattering. Therefore, a correct estimate of the resolution on all spatial coordinates is crucial. The present analysis has been performed on the μ track projected on the 2D wire position and drift time plane corresponding to Collection view, characterized by a better spatial resolution and hit-finding efficiency.

The precision on wire positioning in the T600 detector is better than 0.01 mm [1]. A significant contribution to the uncertainty on the hit position measurement ($\sigma_y$) comes from the $y$ drift coordinate as obtained from a fit of the wire signal shape which is affected by the response of readout electronics and the associated noise.

The single hit uncertainty can be disentangled from the MCS contribution by measuring the dispersion in the drift coordinate with respect to a straight line on a short part of the muon track, where the effect of multiple scattering can be neglected. Sets of three consecutive hits (triplets) have been considered, corresponding to ~0.7 cm segment on average. For each triplet, the distance $\delta_{3P}$ from the middle point to the average position of the two adjacent points is computed; for each event the single point resolution $\sigma_{3P}$ is proportional to the RMS of the $\delta_{3P}$ distribution with a factor √(2/3) obtained from the error propagation. A $\sigma_{3P} \sim 0.7$ mm, consistent with the one obtained from test-pulse signals of similar amplitude, has been determined on average in the stopping muon sample from the distribution of the $\sigma_{3P}$ values as a function of the deposited energy per hit (Figure 10). The obtained $\delta_{3P}$ distance resulted very stable along each muon track, allowing the use of a single value for all hits in an event; comparing the first and second half of each muon, $\delta_{3P}$ results to be stable within ~ 1%..

Another contribution to the drift coordinate uncertainty is given by the relative synchronization of digital electronic boards (32 channels), which is known to be within one sampling time interval ($t_s$ = 400 ns); hence the trigger signal arrives to each board randomly distributed within the sampling time interval. As a result, an additional uncertainty of $\Delta y_{BD} = v_D\ t_s \sim 0.6$ mm, corresponding to a RMS of 0.18 mm, totally correlated between wires in the same board, is introduced.

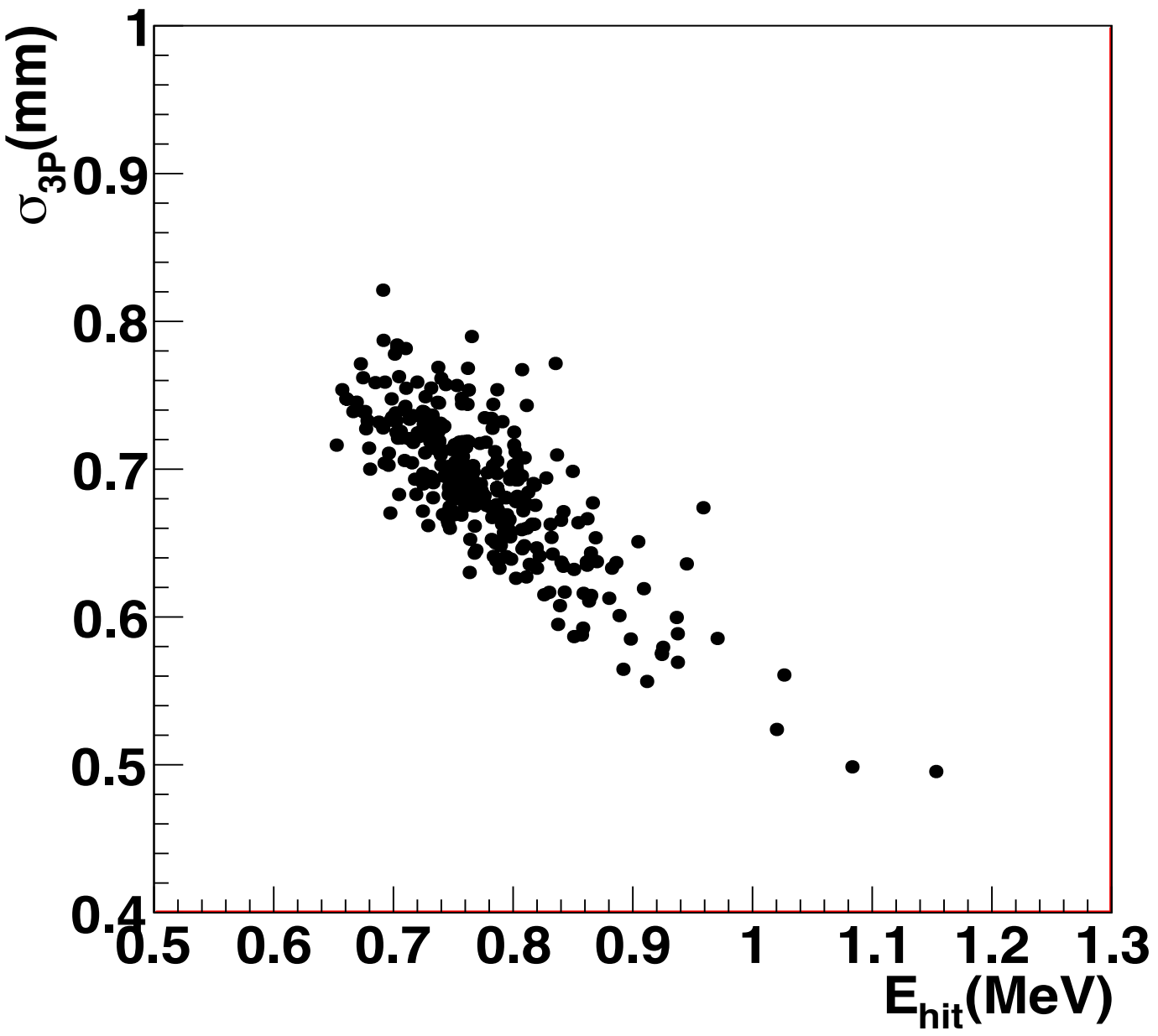

**Figure 10: Scatter-plot of $\sigma_{3P}$ vs. the average deposited energy per hit for each event in the stopping muon sample. An anti-correlation is observed: lower signal height corresponds to a larger error on position.**

## 5. Momentum measurement algorithm

A track-by-track estimation of the muon momentum *p* can be provided by the measurement of the genuine RMS multiple scattering angle $\psi_{MCS}$ over a length *L,* roughly described in LAr within ~1% percent accuracy at $L \sim X_0$ [13]:

$$\psi_{MCS} = \frac{13.6 MeV/c}{\beta p}\sqrt{\frac{L}{X_0}}\left(1+0.038\ln\frac{L}{X_0}\right), \qquad (1)$$

where $X_0$ is the radiation length in liquid Argon and $\beta = v_\mu/c$ . This formula refers to the angle projected on a 2D plane and neglects non-Gaussian tails corresponding to large scatterings, and therefore only approximately describes the Gaussian part of the scattering angle distribution.

The muon track has been initially divided into a number of segments, grouping together hits belonging to two consecutive electronic boards, corresponding to an $L_{seg}$ ~ 19.2 cm and to an average number of $n_{hit}$ = 57 hits per segment for the stopping muon sample. The adopted segmentation represents a compromise between different requirements. Longer segments allow enhancing the physical MCS deflections, that grow as $(L_{seg})^{1/2}$, while reducing the impact of measurement errors on single wires. On the other hand, an adequate segment statistics (> 10) is required to correctly estimate average deflections, even for the shortest muon lengths (2.5 m). Moreover, grouping all hits in a board in the same segment allows reducing the impact of the board synchronization on the momentum resolution.

The muon trajectory within a segment is described by the position of the barycenter of the associated hits, and by its slope defined on the 2D Collection plane. With these parameters, two different definitions of the deflection angle $\theta$ between consecutive segments can be considered. In the first approach (polygonal) $\theta_{poly}$ is defined as the angle between two consecutive pieces of the polygonal line connecting the corresponding barycenter points (Figure 11 left). Alternatively (linear-fit) each segment is fitted independently with a straight line, and the angle $\theta_{lin}$ is defined as the difference between the fitted slopes of two consecutive segments (Figure 11 right).

The RMS deflection angle between two consecutive segments due to multiple scattering can be expressed as:

$$\theta_{MCS} = \frac{13.6 MeV/c}{\beta p_i} \sqrt{\frac{L_{seg}}{X_0 \cos\delta} \frac{w_0}{\cos\delta}} . \tag{2}$$

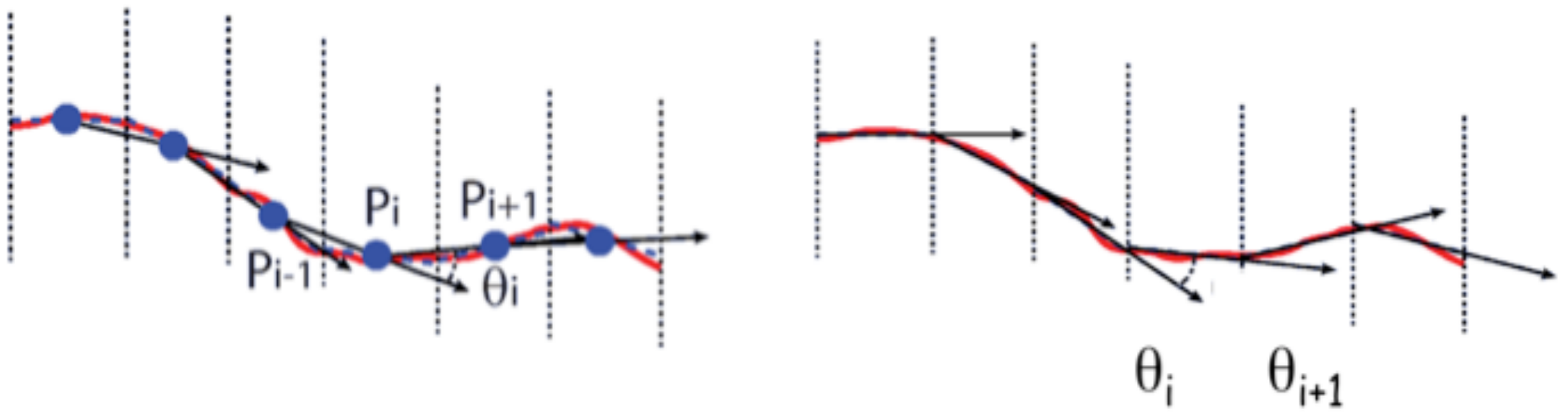

**Figure 11: Scheme of the polygonal (left) and linear-fit (right) deflection angles.**

While equation (1) expresses the local deflection at a point along the muon track, in equation (2) the deflection is averaged over the finite segment length, resulting in a reduction of the MCS effect quantified by the factor $w_0$. For the considered segment length, $w_0$ has been estimated to be $w_{0,poly}$~0.74 and $w_{0,lin}$~0.86 within 1% for the polygonal and the linear-fit respectively, with a numerical model simulating individual Coulomb scatterings on a large number (~$10^6$) of tracks. The momentum $p_i$ refers to the $i$-th segment of the μ track and is computed from the initial momentum $p_0$ assigned to the track accounting for the energy losses along the muon path as measured by the calorimetry. The factors $\cos\delta = L_{2D}/L_{3D}$ and $\sqrt{(\cos\delta)}$ account for the projections on the Collection plane of the 3D MCS angle and the 3D muon length respectively.

In addition to the genuine MCS term $\theta_{MCS}$, the observed deflections also contain independent measurement error contributions $\theta_{3P}$ from the single point space resolution and $\theta_{BD}$ from the board-to-board synchronization. They can be expressed respectively as:

$$\theta_{3P} = \frac{\sigma_{3P}}{L_{seg}} \frac{k_{3P}}{\sqrt{n_{hit}}}, \qquad \theta_{BD} = \frac{\Delta y_{BD}}{L_{seg}} k_{BD}, \tag{3}$$

where $k_{3P}$ is √6 (√24) for the polygonal (linear-fit) angle, the board-to-board term $k_{BD}$ is 1/2 (1/√3) for the polygonal (linear-fit) angle as obtained from error propagation. Both contributions do not depend on momentum; $\theta_{BD}$ ~2 mrad is dominant over $\theta_{3P}$ ~1 mrad. The MCS term obviously depends on $p_i$: for 2 GeV/c muon momentum $\theta_{MCS}$ ~2 mrad is comparable with $\theta_{3P}$ and $\theta_{BD}$.

In order to compare the observed deflections on a μ track with the expectations for a given momentum $p$, a $\chi^2$-like function $C_2$ has been constructed as:

$$C_2(p) = {}^tVC(p)^{-1}V = {}^tV(C_{MCS}(p) + C_{3P} + C_{BD})^{-1}V \tag{4}$$

where $V$ is the vector which contains all $2n_{seg}$-$3$ observed scattering angles, $n_{seg}$-$2$ computed with the polygonal definition and $n_{seg}$-$1$ with the linear fit. The covariance matrix $C$ expresses the expected deflections and their mutual correlations; the diagonal terms correspond to the sum of squares of the deflection angles defined in (2) and (3):

$$C_{ii}(p) = \theta^2_{MCS}(p_i) + \theta^2_{3P} + \theta^2_{BD} \tag{5}$$

The momentum-dependent MCS covariance matrix $C_{MCS}(p)$ can be expressed as follows:

$$C_{MCS}(p) = \begin{pmatrix} C^{poly}_{MCS}(p) & C^{mix}_{MCS}(p) \\ {}^tC^{mix}_{MCS}(p) & C^{lin}_{MCS}(p) \end{pmatrix} \tag{6}$$

where the sub-matrices $C^{poly}{}_{MCS}(p)$ and $C^{lin}{}_{MCS}(p)$ refer to the polygonal and linear-fit angles respectively, and $C^{mix}{}_{MCS}(p)$ accounts for their cross-correlations introduced by the MCS. The detailed expressions for all these sub-matrices can be found in Appendix.

Measurement errors, both from the single-point resolution $\sigma_{3P}$ and the board-to-board synchronization $\sigma_{BD}$, are defined within a single segment excluding any correlation between the different segments. The only covariance terms derive from the definition of the deflection angles as difference between neighboring segments: the same segment appears in 3 (2) deflection angles in the polygonal (linear-fit) case. Moreover, the single-point measurement errors for the polygonal and the linear-fit procedures are fully uncorrelated between each other; the correlation between the board-to-board ones, also vanishing in first-order approximation, has also been neglected.

The above analysis involves Gaussian contribution to the various errors, while the physical MCS deflection angle distribution also exhibits significant non-Gaussian tails which can affect the MCS momentum measurement. Therefore, single scattering terms contributing more than 3σ to the $C_2(p)$ function are excluded from the computation.

The determination of the μ momentum is based on the analysis of $C_2$ function of the unknown initial track momentum p. The best estimate $p_{MCS}$ - given the observed deflections - is obtained assuming $C_2(p_{MCS}) = (2\ n_{seg}\ -3)$, i.e. the RMS of the observed deflection angles matches on average the value expected from the instrumental and MCS contributions. The $C_2$ function is well described around this best value by:

$$C_2/(2n_{seg}-3)=\frac{1}{\alpha+\beta/p^2}, \quad (7)$$

where $\alpha$ and $\beta/p^2$ refer to the instrumental and MCS contributions to the observed deflection respectively (see Figure 12).

The relative uncertainty of the $C_2$ function is $\Delta C_2/C_2 = \sqrt{(2/(2\, n_{seg} -3))}$ since for the true μ momentum the $C_2$ function is distributed as a $\chi^2$ with $(2\, n_{seg} -3)$ degrees of freedom. The associated relative resolution on the momentum estimation is then:

$$\frac{\sigma_p}{p}=\frac{1}{\sqrt{2(2n-3)}}(1+\frac{\alpha p^2}{\beta}) \quad (8)$$

where the $2(2n_{seg} -3)^{-1/2}$ term, which accounts for the statistical uncertainty in the $C_2$ distribution, corresponds to the lower limit of the resolution as obtained at very low $p$ values. The term proportional to $p^2$ contains the momentum dependency; as expected the resolution gets worse as $p$ increases. Both these uncertainties are graphically summarized in the red bands in Figure 12.

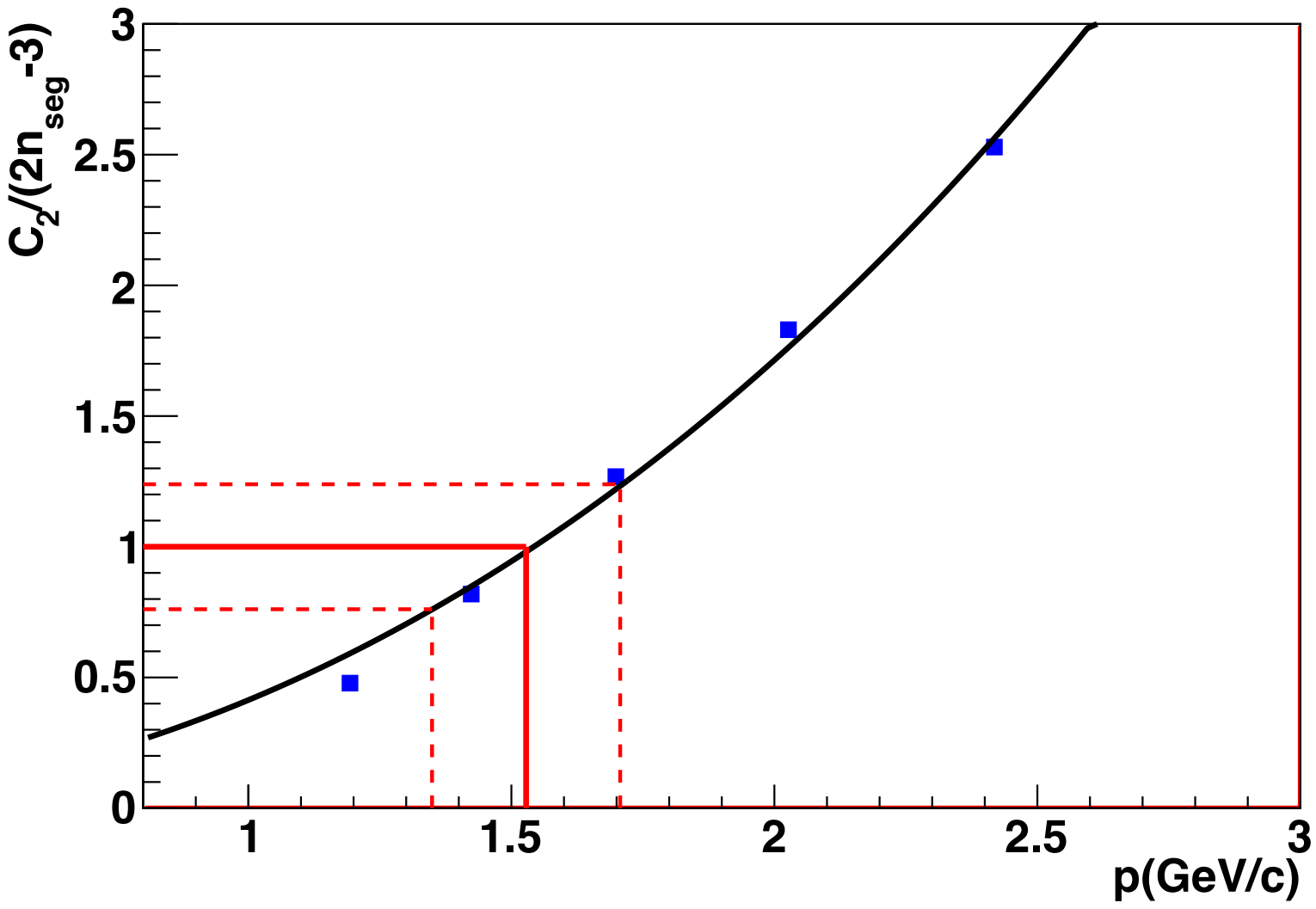

**Figure 12: Example of fit to the $C_2(p)$ function for a stopping μ. The calorimetric value of momentum is 1.6 GeV/c. A 4 m track length is used in the analysis corresponding to $n_{seg}$ = 19. The fitted values of α and β are ~0.03 and 2.4 (GeV/c)$^2$ respectively. The resulting momentum is $p_{MCS}$~ 1.5 GeV/c. The dashed red lines refer to the 1σ allowed range.**

## 6. Momentum measurement on the CNGS stopping muons

The calorimetric measurement of the stopping muon momentum can provide an adequate direct benchmark of the MCS momentum estimation. The algorithm has been initially applied to a sample of 1000 simulated horizontal stopping muon events with a flat energy spectrum from 1 to 5 GeV. The simulation, performed with the FLUKA package [14] , included a detailed model of the muon transport within the T600 volume, with detailed simulation of energy losses by ionization, delta ray production, electromagnetic and hadronic interactions. All known experimental effects such as recombination and signal attenuation due to finite electron lifetime are taken into account. The resulting energy deposition is registered in cells with a fine space and time (50 ns) granularity in order to precisely model the tiny deflections due to MCS. The corresponding charge is convoluted with the known response function of the ICARUS-T600 front-end electronics; noise is modeled according to the observed noise conditions at LNGS.The comparison between MCS and calorimetric estimations for these events (Figure 13) showed a good agreement on average, with no evidence of relevant bias.

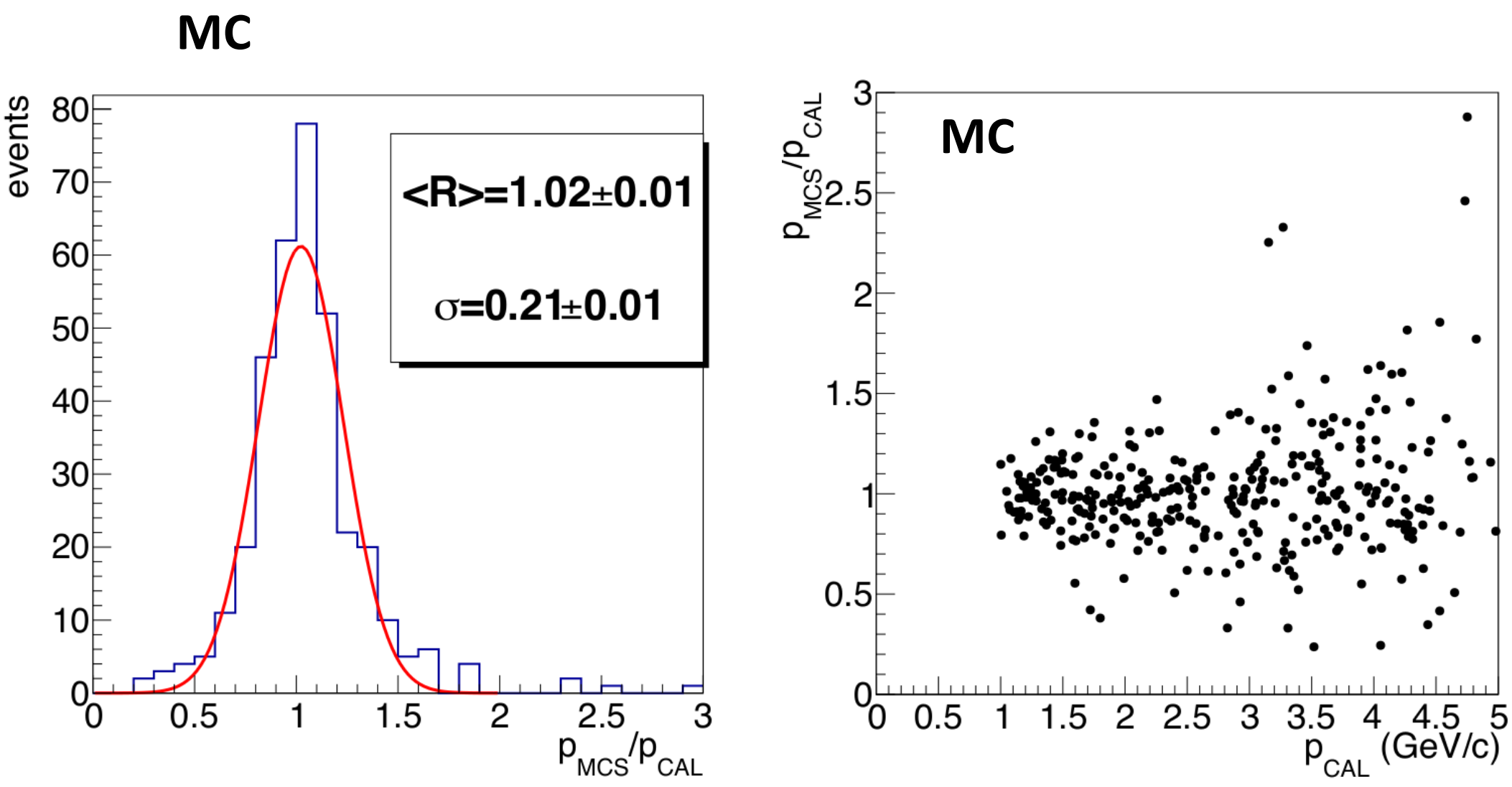

**Figure 13:Distribution of $p_{MCS}/p_{CAL}$ ratio centered at ~1 (left) and $p_{MCS}/p_{CAL}$ as a function of $p_{CAL}$ (right) for the MC stopping muons ($L_{\mu}$= 4 m). While the resolution varies depending on muon momentum, an average resolution ~21% can be obtained from a Gaussian fit in this energy range.**

However, distortions $\delta y$ on the drift coordinate due to the small local deviation of the drift electric field and drift velocity from their 500 V/cm and $v_D$ ~ 1.6 mm/μs nominal values, generated by the observed cathode non-planarity in the East module, have to be properly taken into account. A local few percent change of $v_D$ would introduce an additional track deviation θ ~ 1 mrad mimicking the MCS. Such effect is expected to bias the $p_{MCS}$ measurement at high values of muon momenta where the genuine deflection angle associated to MCS is small, resulting in an underestimation of the actual momentum value.

The 3D map of the local $\delta y$ variations, as obtained from the measured cathode distortions, has been calculated on the full active TPC volume of the East module and inserted in the stopping muon simulation Monte Carlo code. About 1000 stopping muon events have been produced for each of three 50 cm width detector slices of the drift path to evaluate the impact on the momentum measurement by the multiple scattering at the different distances from the cathode. As a result an underestimation of $p_{MCS}$ w.r.t. $p_{CAL}$ as a function of the calorimetric momentum measurement has been observed in the Monte Carlo, especially for stopping μs travelling near the cathode (Figure 14).

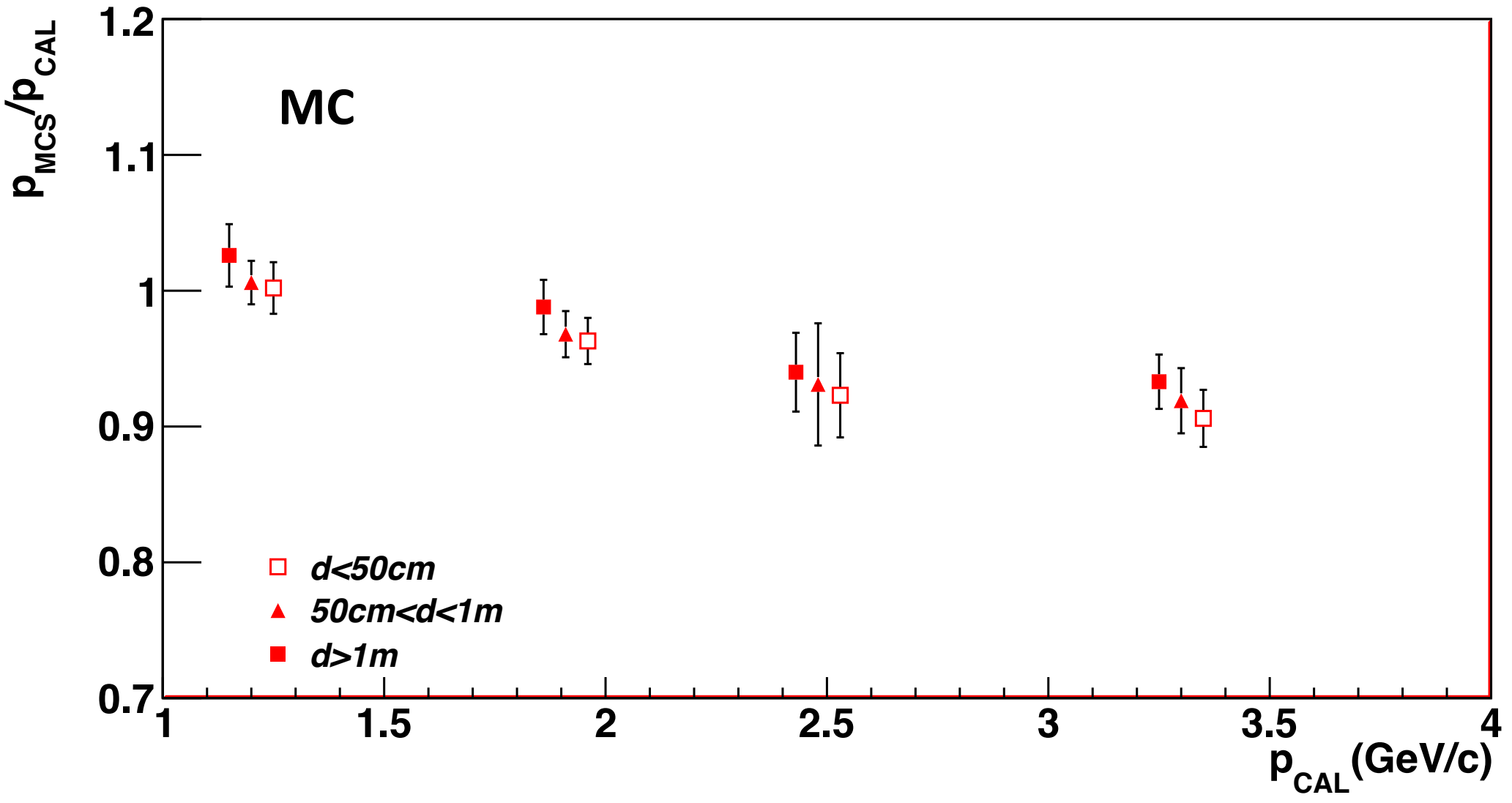

**Figure 14: MC calculation of μ momentum measurement by MCS in the East module with included the effective electric drift field as determined from the cathode planarity measurements. Muons have been grouped in 3 bins of distance from the cathode, ~50 cm each, spanning the full drift path of the TPC. The average energies in each bin (on the horizontal axis) are shifted by ±50 MeV to enhance visibility.**

Then the momentum measurement algorithm has been applied to the selected CNGS stopping μs as recorded in both East and West modules, and the resulting $p_{MCS}$ has been compared event by event to the corresponding calorimetric measure $p_{CAL}$ (Figure 15 and Figure 16). The observed correlation between $p_{MCS}$ and $p_{CAL}$ demonstrates the reliability of the proposed method in the considered momentum range. The $p_{MCS}/p_{CAL}$ distribution deviates from a Gaussian and presents significant tails, associated with events with large Coulomb scatterings, with a full RMS value of ~18%. Since the error on $p_{CAL}$ is much smaller than on $p_{MCS}$, the width of this distribution provides a good average estimate of the uncertainty ($\sigma(p_{MCS})/p_{MCS}$) on the MCS momentum measurement. The RMS width of the distribution bulk can be more accurately inferred from the FWHM, resulting ~14%; an approximate Gaussian fit to the $p_{MCS}/p_{CAL}$ distribution results in σ~16%. The resolution is slightly better than for the Monte Carlo sample because of the lower average momentum of the data.

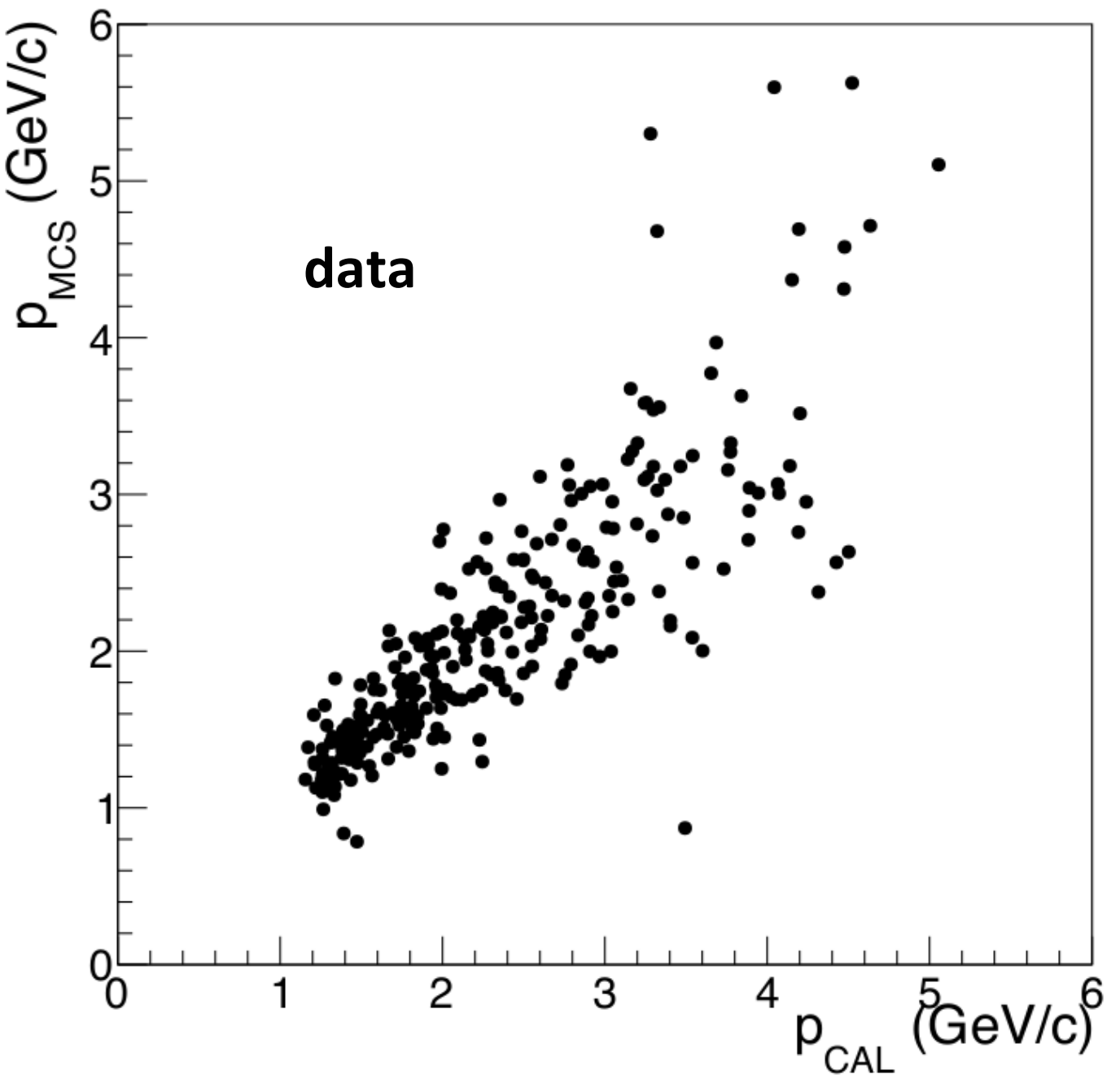

**Figure 15: Scatter plot of $p_{MCS}$ vs. $p_{CAL}$ for the CNGS stopping μ data. The first 4 meters of track were used for the momentum determination excluding the last meter before range-out (only tracks exceeding 5 m length have been considered, see Paragraph 3).**

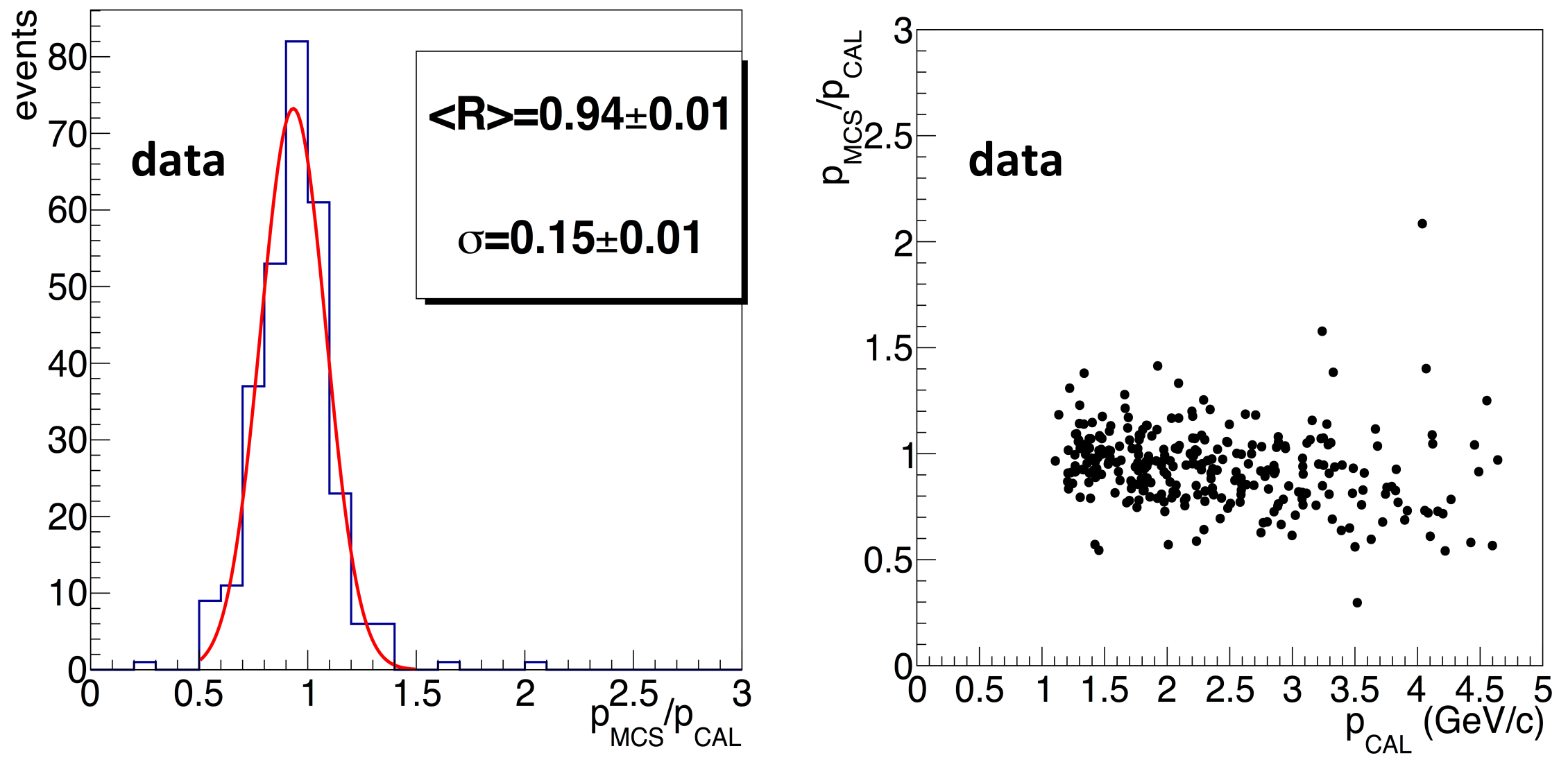

**Figure 16: Distribution of $p_{MCS}/p_{CAL}$ ratio for the stopping muon sample (left) and $p_{MCS}/p_{CAL}$ ratio vs. $p_{CAL}$ (right) for $L_{\mu}$= 4 m. The FWHM width of the $p_{MCS}/p_{CAL}$ distribution is ~0.33.**

The $p_{MCS}/p_{CAL}$ ratio appears to slightly decrease as the muon momentum grows as observed in the MC analysis of the East module when the effects of the cathode non-planarity have been considered. At low muon momenta it is consistent with unity, while a uniformly decreasing trend appears at higher values, reaching a ~15-20 % underestimation above 3.5

GeV/c. Moreover data exhibit also a dependency of the $p_{MCS}/p_{CAL}$ ratio on the track distance from the cathode (Figure 17) as predicted in the MC events. Restricting the data analysis to the event sample recorded in the East module, the underestimation introduced by the cathode effect is slightly reduced to ~15% at highest muon momenta, resulting in a closer agreement with the corresponding MC calculation.

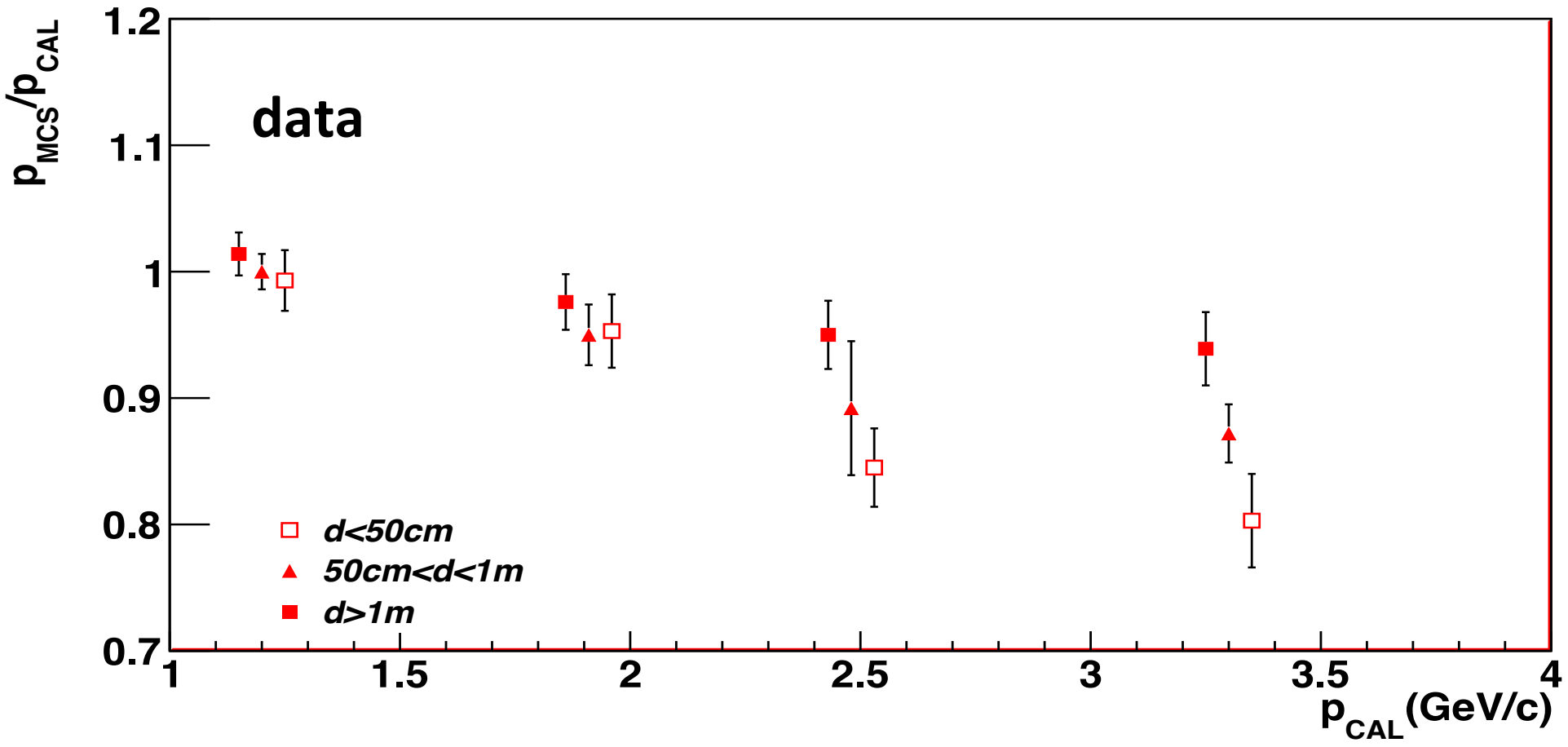

**Figure 17: Dependency of $p_{MCS}/p_{CAL}$ ratio on $p_{CAL}$ (for $L_{\mu}$=4 m) for the full stopping muon sample in both modules; data have been grouped in 3 bins of distance from the cathode, ~50 cm each, spanning the full drift path of the TPC. The average energies in each bin (on the horizontal axis) are shifted by ±50 MeV to enhance visibility.**

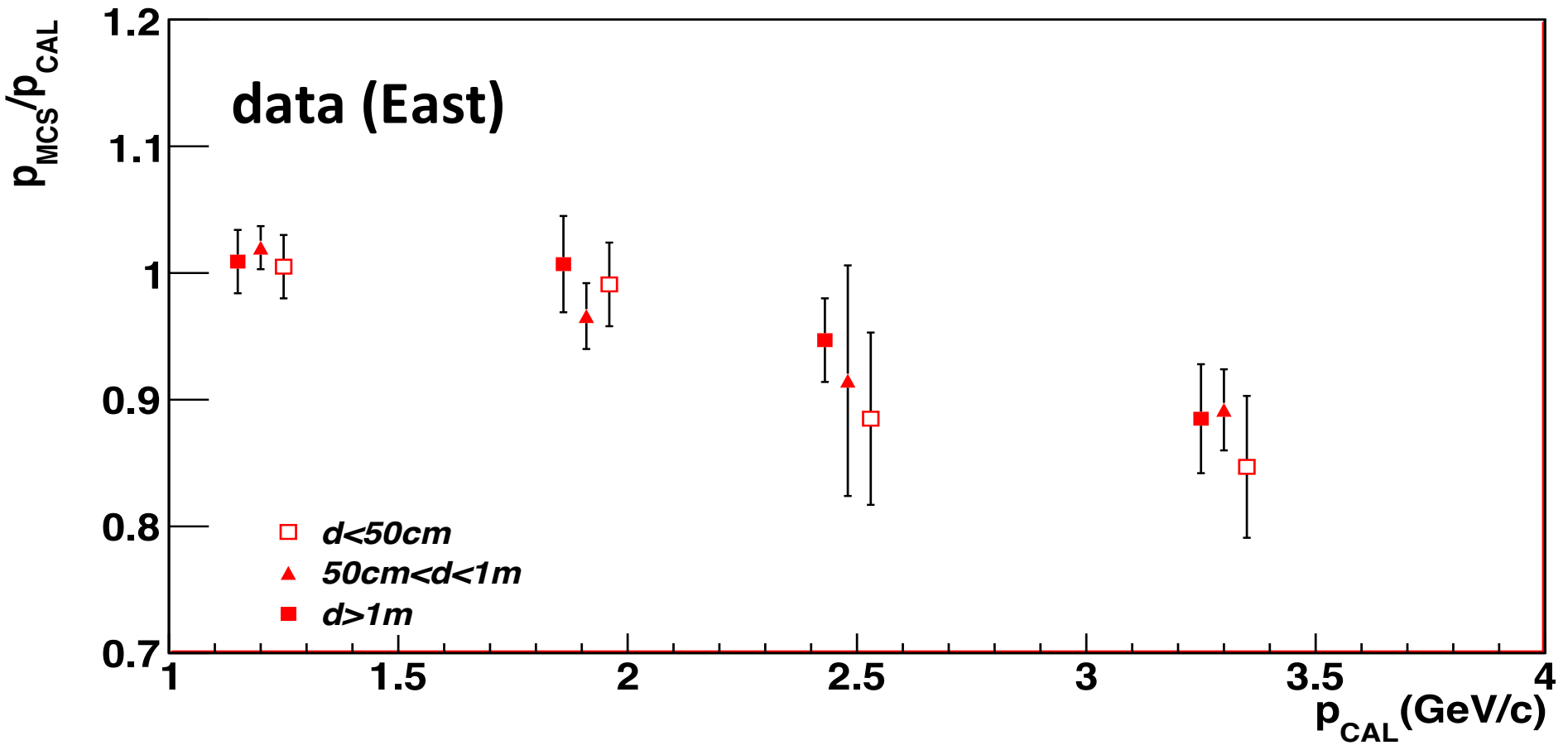

**Figure 18: Dependency of $p_{MCS}/p_{CAL}$ ratio on $p_{CAL}$ (for $L_{\mu}$=4 m) for muons in the East module only; data have been grouped depending on the distance from the cathode. The average energies in each bin (on the horizontal axis) are shifted by ±50 MeV to enhance visibility.**

The dependency of the momentum resolution $\sigma_p/p$ both on true momentum $p_{CAL}$ and the track length used in the analysis is shown in Figure 19. As expected, $\sigma_p/p$ strongly improves for larger track lengths due to the increase in the segment statistics used in the momentum calculation.

The resolution also strongly improves at lower values of muon momentum. In the region between 1 and 1.5 GeV/c, which is of interest to next SBN and long baseline neutrino experiments, a value close to 11% can be achieved for $L_\mu$ = 4m.

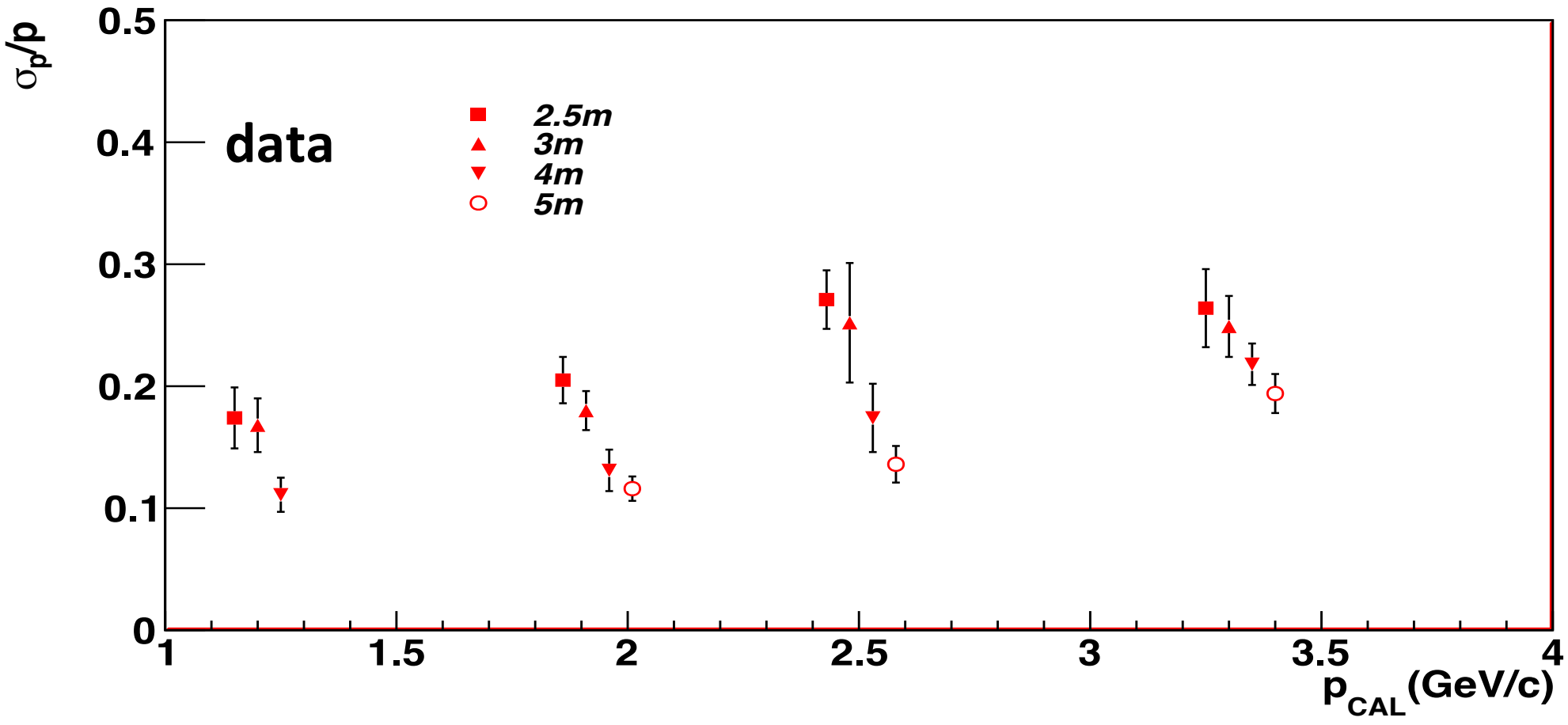

**Figure 19: Dependency of momentum resolution on calorimetric muon momentum for different values of the used muon track length $L_\mu$. The average energies in each bin (on the horizontal axis) are shifted by 50 MeV to enhance visibility.**

The difference of the measured $p_{MCS}$ from the calorimetric measurement has been compared with the estimated error $\sigma_p$, as defined in equation (8), by the pull $(p_{MCS}-p_{CAL})/\sigma_p$. The width of the pull distribution (Figure 20) is compatible with unity, confirming that this error parameterization correctly describes, on average, the data dispersion. The asymmetry of the curve is due to the small momentum underestimation, already discussed in Figure 16. The present algorithm can provide a measurement of muon momentum with an average resolution ~16% up to ~3.5 GeV/c and therefore be used for a full reconstruction of $\nu_\mu$CC events in the forthcoming low-energy neutrino experiments [3][[4].

A local point-by-point correction to track distortions due to the cathode deviations from planarity would require a too complex and extended calculation of the electric drift field including in Comsol all the mechanical details of the panels composing the central cathode and the surrounding field cage electrodes. Moreover the measurement of the cathode deviations in the East Module has been performed with empty module at room temperature, i.e. under very different conditions compared to the Gran Sasso data taking and after its transport from Gran Sasso to CERN. As a first approximation, an average event-by-event correction to momentum measurement by MCS for the cathode deformation can be performed by applying the ratio $(p_{CAL}/p_{MCS})_{MC}$ as obtained in the Monte Carlo analysis as a function of momentum and distance from the cathode (as seen in Figure 14). As a result the observed underestimation of $p_{MCS}$ to $p_{CAL}$ was strongly reduced within the quoted errors (Figure 21) from ~15 % to ~5% at 3.5 GeV/c. The corresponding momentum resolution averaged on the East module stopping μ sample was improved to 14 %.

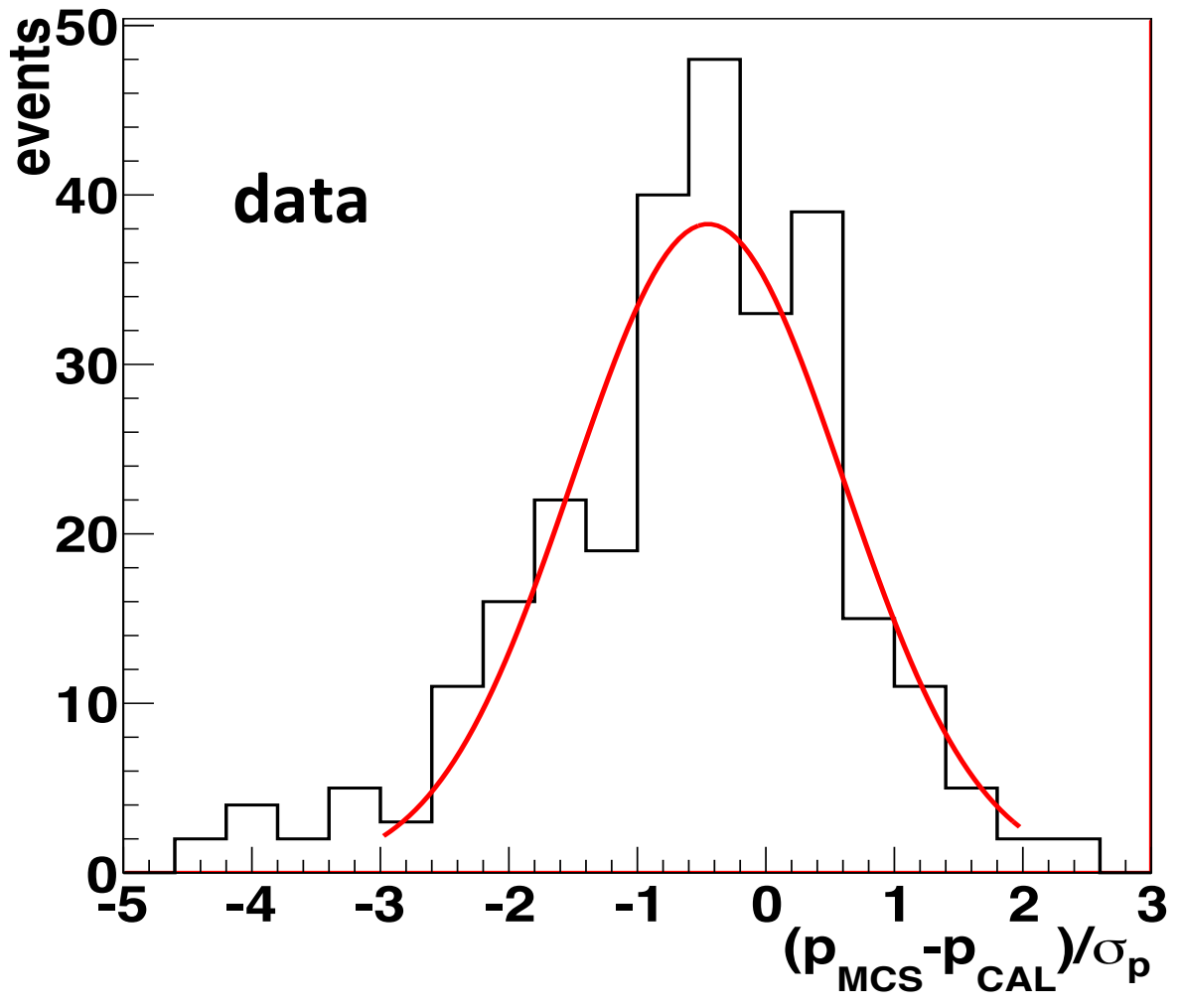

**Figure 20: Distribution of the "pull" quantity $(p_{MCS}-p_{CAL})/\sigma_p$ on the whole stopping muon sample. The first 4 meters of muon track were used. The RMS of a Gaussian fit of the bulk of the distribution is 1.05±0.07, while the peak is at -0.44±0.07.**

An improved TPC cathode planarity combined with a full synchronization of electronic read-out boards, allowing a better optimization of segment length, should permit to extend the measurement to higher muon momenta well covering the value range required in the forthcoming short and long base-line neutrino experiments, while also improving the resolution by ~20%.

At the shallow-depth location of the SBN experiment, the detector will be exposed to a much larger cosmic ray flux (by ~6 orders of magnitude) than in the LNGS underground laboratory. The space charge effects due to the accumulation of positive ions by cosmic rays are expected to modify the electric field ([16]) and the corresponding drift velocity, which are reduced close to the anode and increased near the cathode. This results in a distortion of the reconstructed tracks, possibly leading to an additional systematic error on the MCS muon measurement. This effect is expected to grow as the cube of the drift distance, and inversely proportional to the electric field.

In the ICARUS-T600 case (~500 V/cm field) dedicated studies were performed on cosmic muons recorded in a surface test-run ([17]), crossing the full drift distance of a single TPC (~1.5 m drift distance) at an angle of ~60° w.r.t. the drift direction. These muons exhibit a slightly parabolic distortion with a sagitta which is on average ~3 mm; the same study performed on ICARUS-T600 at the LNGS underground location showed a negligible effect.

The distortion due to space charge, expected at SBN, will then be slightly smaller than that measured in the surface test-run due to the 3 m concrete overburden that will be placed over the detector at FNAL, thus reducing the cosmic ray flux by ~30%. These distortions will be however smaller that that due to the residual cathode non-planarity described above. The space charge effect is therefore not expected to limit the measurement performance at the ~ GeV energy range of the Booster beam neutrinos.

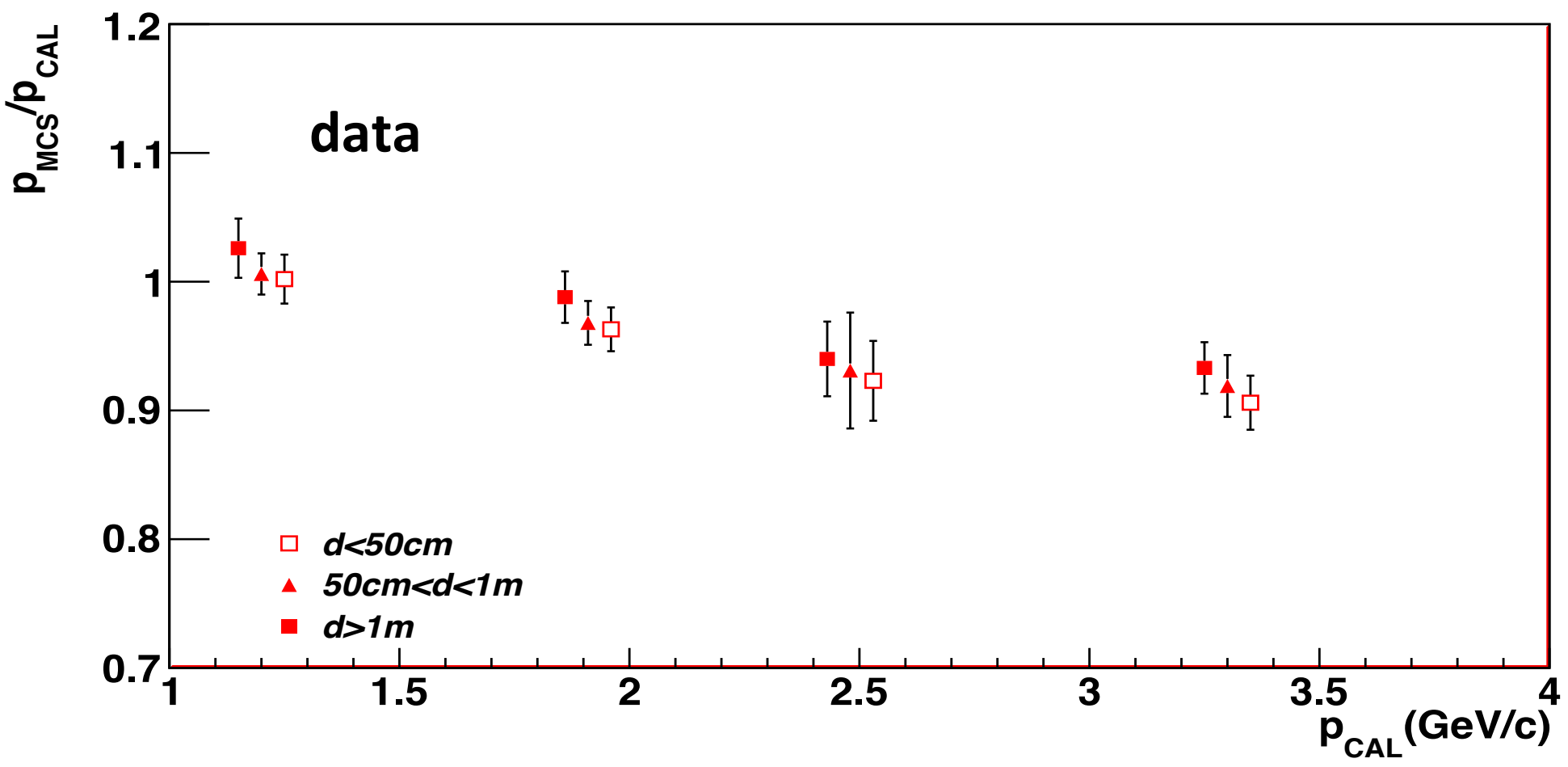

**Figure 21: Dependency of $p_{MCS}/p_{CAL}$ ratio on $p_{CAL}$ (for $L_\mu$=4m) for CNGS muons stopping in the East module (3 bins of muon distance from the cathode) after the average correction for the cathode deviations from planarity as determined from the corresponding MC calculation. The average energies in each bin (on the horizontal axis) are shifted by ±50 MeV to enhance visibility.**

## 7. Conclusions

The measurement of muon momentum by means of multiple Coulomb scattering is a crucial ingredient to the reconstruction of $\nu_\mu$CC events in LAr-TPCs in the absence of a magnetic field. A muon momentum measurement algorithm has been developed for the ICARUS-T600 experiment at the INFN-LNGS underground laboratory. It has been validated on a sample of ~500 recorded muon tracks produced in the CNGS beam neutrino interactions in the rock surrounding the detector and stopping in the LAr-TPC.

The present analysis extends the study with LAr-TPC performed in the sub-GeV energy range with cosmic muons up to 5 GeV, covering an energy region which is relevant for the next SBN and long base-line neutrino experiments. As a result the muon momentum can be measured by using the multiple Coulomb scattering approach with an average resolution of ~16% which is expected to improve to ~14% if the observed cathode plane non-planarity is taken into account. These observed effects are expected to be strongly reduced by the cathode flattening procedure as currently applied during the T600 overhauling at CERN.

## Acknowledgements

The ICARUS Collaboration acknowledges the fundamental contribution of INFN to the construction and operation of the experiment. In particular, the authors are indebted to the LNGS Laboratory for the continuous support to the experiment. The Polish groups acknowledge the support of the National Science Center, Harmonia (2012/04/M/ST2/00775). Finally, we thank CERN, in particular the CNGS staff, for the successful operation of the neutrino beam facility.

## Appendix: Covariance matrices of deflection angles

The covariance sub-matrix describing the MCS contribution to the deflection angles computed according to the polygonal approach has the 5-diagonal symmetrical form:

$$C_{MCS}^{poly}(p)=\begin{pmatrix} \theta_1^2 & \alpha_+^p\theta_1\theta_2 & \alpha_{++}^p\theta_1\theta_3 & 0 & \cdots \\ \alpha_+^p\theta_1\theta_2 & \theta_2^2 & \alpha_+^p\theta_2\theta_3 & \alpha_{++}^p\theta_2\theta_4 & 0 \\ \alpha_{++}^p\theta_1\theta_3 & \alpha_+^p\theta_2\theta_3 & \theta_3^2 & \alpha_+^p\theta_3\theta_4 & \alpha_{++}^p\theta_3\theta_5 \\ 0 & \alpha_{++}^p\theta_2\theta_4 & \alpha_+^p\theta_3\theta_4 & \theta_4^2 & \alpha_+^p\theta_4\theta_5 \\ \vdots & 0 & \alpha_{++}^p\theta_3\theta_5 & \alpha_+^p\theta_4\theta_5 & \theta_5^2 \end{pmatrix} \qquad (9)$$

where $\theta_i=\theta_{MCS}(p_i)$ is the polygonal-line scattering angle at the i-th segment (see Equation 2, with $w_0=w_{0,poly}$), and the constants $\alpha^p{}_+$ and $\alpha^p{}_{++}$ were computed numerically with the same model used for the estimation of $w_0$.

The sub-matrix involving the linear-fit terms has a 3-diagonal symmetrical form since only two segments are involved in the deflection angle definition:

$$C_{MCS}^{lin}(p)=\begin{pmatrix} \theta_{0,1}^2 & \alpha_+^l\theta_{0,1}\theta_{1,2} & 0 & \cdots & \cdots \\ \alpha_+^l\theta_{0,1}\theta_{1,2} & \theta_{1,2}^2 & \alpha_+^l\theta_{1,2}\theta_{2,3} & 0 & \cdots \\ 0 & \alpha_+^l\theta_{1,2}\theta_{2,3} & \theta_{2,3}^2 & \alpha_+^l\theta_{2,3}\theta_{3,4} & 0 \\ \vdots & 0 & \alpha_+^l\theta_{2,3}\theta_{3,4} & \theta_{3,4}^2 & \ddots \\ \vdots & \vdots & 0 & \ddots & \ddots \end{pmatrix} \qquad (10)$$

where, similarly, $\theta_{i,i+1}=\theta_{MCS}(p_{i,i+1})$ is the linear-fit scattering angle between the i-th and (i+1)-th segments (see Equation 2, with $w_0=w_{0,lin}$)., $p_{i,j}$ being the average momentum between two adjacent segments ($p_{i,i+1} = (p_i+p_{i+1})/2$).

The (n-2)×(n-1) mixing sub-matrix referring to the correlation between polygonal and linear-fit angles (adopting similar notations as in (9) and (10) for the scattering angles) is:

$$C_{MCS}^{mix}(p)=\begin{pmatrix} \beta_+\theta_1\theta_{0,1} & \beta_+\theta_1\theta_{1,2} & \beta_{++}\theta_1\theta_{2,3} & 0 & \cdots \\ \beta_{++}\theta_2\theta_{0,1} & \beta_+\theta_2\theta_{1,2} & \beta_+\theta_2\theta_{2,3} & \beta_{++}\theta_2\theta_{3,4} & 0 \\ 0 & \beta_{++}\theta_3\theta_{1,2} & \beta_+\theta_3\theta_{2,3} & \beta_+\theta_3\theta_{3,4} & \beta_{++}\theta_3\theta_{4,5} \\ \vdots & 0 & \beta_{++}\theta_4\theta_{2,3} & \beta_+\theta_4\theta_{3,4} & \beta_+\theta_4\theta_{4,5} \\ \vdots & \vdots & 0 & \ddots & \ddots \end{pmatrix} \qquad (11)$$

Each polygonal term is correlated to 4 linear-fit terms and vice-versa, because the polygonal term involves 3 segments (i-1,i,i+1) while a linear term involves only two (i,i+1). The numerically computed $\beta_+$ and $\beta_{++}$ coefficients parameterize the cases when the polygonal and linear-fit terms share two and one common segments respectively.

Measurement errors, both from the single-point resolution $\sigma_{3P}$ and the board-to-board misalignment $\sigma_{BD}$, are defined within a single segment excluding any correlation between the different segments, as explained in the text. Therefore, the measurement error covariance matrix can be expressed as follows: the factors $\theta_{3P}^2$ and $\theta_{BD}^2$, that assume different numerical values in the polygonal and linear-fit cases, express the independent angular uncertainties due to the single-point resolution and board-to board alignment respectively. They are multiplied by a matrix that expresses the correlations between deflections at successive segments, and has the same form for both uncertainty sources.

$$C_{meas} = C_{3P} + C_{BD} = \begin{pmatrix} C_{meas}^{poly} & 0 \\ 0 & C_{meas}^{lin} \end{pmatrix} \tag{12}$$

$$C_{meas}^{poly} = \left[\theta_{3P}^2 + \theta_{BD}^2\right] \begin{pmatrix} 1 & -\frac{2}{3} & \frac{1}{6} & 0 & \cdots \\ -\frac{2}{3} & 1 & -\frac{2}{3} & \cdots & \cdots \\ \frac{1}{6} & -\frac{2}{3} & 1 & \cdots & \cdots \\ 0 & \vdots & \vdots & \ddots & \ddots \\ \vdots & \vdots & \vdots & \ddots & \ddots \end{pmatrix} \quad C_{meas}^{lin} = \left[\theta_{3P}^2 + \theta_{BD}^2\right] \begin{pmatrix} 1 & -\frac{1}{2} & 0 & 0 & \cdots \\ -\frac{1}{2} & 1 & -\frac{1}{2} & \cdots & \cdots \\ 0 & -\frac{1}{2} & 1 & \cdots & \cdots \\ 0 & \vdots & \vdots & \ddots & \ddots \\ \vdots & \vdots & \vdots & \ddots & \ddots \end{pmatrix} \tag{13}$$